\documentclass[12pt,preprint]{emulateapj}
\usepackage{lscape}
\usepackage{color}
\usepackage{url}

\newcommand{\rosat}{{\sl ROSAT}}
\newcommand{\galex}{{\sl GALEX}}

\begin{document}

\title{A Stellar Census of the Tucana-Horologium Moving Group}

\author{
Adam L. Kraus\altaffilmark{1,2,3},
Evgenya L. Shkolnik\altaffilmark{4},
Katelyn N. Allers\altaffilmark{5},
Michael C. Liu\altaffilmark{6}
}

\altaffiltext{1}{Dept of Astronomy, The University of Texas at Austin, Austin, TX 78712, USA}
\altaffiltext{2}{Harvard-Smithsonian Center for Astrophysics, 60 Garden St., Cambridge, MA 02138, USA}
\altaffiltext{3}{Clay Fellow}
\altaffiltext{4}{Lowell Observatory, 1400 West Mars Hill Road, Flagstaff, AZ, 86001, USA}
\altaffiltext{5}{Department of Physics and Astronomy, Bucknell University, Lewisburg, PA 17837, USA}
\altaffiltext{6}{Institute for Astronomy, University of Hawaii at Manoa, 2680 Woodlawn Dr., Honolulu, HI 96822, USA}

\begin{abstract}

We report the selection and spectroscopic confirmation of 129 new late-type (SpT$=$K3--M6) members of the Tucana-Horologium moving group, a nearby ($d \sim 40$ pc), young ($\tau \sim 40$ Myr) population of comoving stars. We also report observations for 13 of the 17 known Tuc-Hor members in this spectral type range,  and that 62 additional candidates are likely to be unassociated field stars; the confirmation frequency for new candidates is therefore $129/191 = 67\%$. We have used radial velocities, H$\alpha$ emission, and Li$_{6708}$ absorption to distinguish between contaminants and bona fide members. Our expanded census of Tuc-Hor increases the known population by a factor of $\sim$3 in total and by a factor of $\sim$8 for members with SpT$\ge$K3, but even so, the K-M dwarf population of Tuc-Hor is still markedly incomplete. Our expanded census allows for a much more detailed study of Tuc-Hor than was previously feasible. The spatial distribution of members appears to trace a two-dimensional sheet, with a broad distribution in $X$ and $Y$, but a very narrow distribution ($\pm$5 pc) in $Z$. The corresponding velocity distribution is very small, with a scatter of $\pm$1.1 km/s about the mean $UVW$ velocity for stars spanning the entire 50 pc extent of Tuc-Hor. We also show that the isochronal age ($\tau \sim 20$--30 Myr) and the lithium depletion boundary age ($\tau \sim 40$ Myr) disagree, following the trend in other pre-main sequence populations for isochrones to yield systematically younger ages. The H$\alpha$ emission line strength follows a trend of increasing equivalent width with later spectral type, as is seen for young clusters. We find that moving group members have been depleted of measurable lithium for spectral types of K7.0--M4.5. None of our targets have significant infrared excesses in the WISE W3 band, yielding an upper limit on warm debris disks of $F < 0.7\%$. Finally, our purely kinematic and color-magnitude selection procedure allows us to test the efficiency and completeness for activity-based selection of young stars. We find that 60\% of K-M dwarfs in Tuc-Hor do not have ROSAT counterparts and would have been omitted in X-ray selected samples. In contrast, GALEX UV-selected samples using a previously suggested criterion for youth achieve completeness of 77\% and purity of 78\%, and we suggest new SpT-dependent selection criteria that will yield $>$95\% completeness for $\tau \sim 40$ Myr populations with GALEX data available.
\end{abstract}


\section{Introduction}

Over the past 20 years, co-moving associations of young stars ($\tau \la 100$ Myr) have been identified among the nearby field population \citep[][]{Kastner:1997kx,Webb:1999tw,Mamajek:1999ul,Torres:2000qy,Zuckerman:2000uq,Zuckerman:2004fj}. These moving groups represent the dispersed remnants of coeval stellar populations \citep[e.g.,][]{Weinberger:2012qy} that apparently formed in the same star-forming region, and might be older analogs to unbound associations like Taurus-Auriga and Upper Scorpius \citep[][]{Kraus:2008fr}. Most of these populations are associated with well-known isolated classical T Tauri stars (such as the TW Hya Association, or TWA) or debris disk hosts (such as the $\beta$ Pic moving group, or BPMG), an association which provided the first indication that they were post-T Tauri associations. Surveys to identify active young stars within the solar neighborhood ($d \la 50$ pc) have subsequently identified several additional populations, including the AB Dor, Carina-Near, Hercules-Lyra, and Tucana-Horologium associations \citep[][]{Zuckerman:2004fj,Zuckerman:2006fk,Torres:2008lr,Eisenbeiss:2013qy}.

Young moving groups ($\tau \sim 8$--300 Myr) provide a critical link between star-forming regions (which can be recognized by the presence of molecular cloud material and the preponderance of protoplanetary disk hosts) and the old field population. The close proximity of these young moving groups makes them especially advantageous for the study of circumstellar processes that depend on angular resolution (such as multiple star formation; \citealt[][]{Brandeker:2003wd,Brandeker:2006cr,Evans:2012wd}) and searches for extrasolar planets \citep[][]{Marois:2008zt,Lagrange:2009fc}. The low distances also result in additional sensitivity for flux-limited studies of disks \citep[][]{Low:2005lq,Bouwman:2006yq,Plavchan:2009dp} and the (sub)stellar mass function \citep[][]{Gizis:2002lr,Lyo:2006fk,Murphy:2010lr,Shkolnik:2011qf}. Finally, these stellar populations record a key epoch of planet formation, representing the end of giant planet formation and the onset of terrestrial planet assembly.

The Tucana-Horologium moving group (hereafter Tuc-Hor) is a particularly intriguing stellar population. Its members were first identified separately as the Tucana association and the Horologium association \citep[][]{Torres:2000qy,Zuckerman:2000uq}, but were subsequently recognized to represent a single comoving population with an age of $\tau \sim 30$ Myr. Tuc-Hor is host to at least 12 BAF-type stars \citep[e.g.,][]{Zuckerman:2004fj,Torres:2008lr}, similar in size to the BPMG (also with 12 BAF-type stars) and much larger than the TWA (with a single BAF-type member). Tuc-Hor is likely one of the largest young stellar populations within $d < 100$ pc, making it a robust site for measuring population statistics (the IMF, multiplicity properties, disk frequencies, and activity rates). As an ``intermediate age'' moving group, Tuc-Hor represents a key calibration for age indicators like H$\alpha$ emission, UV and X-ray excesses, rotational velocities, and age-dependent spectral features like $Li_{6708}$, and $Na_{8189}$. If these indicators can be robustly calibrated for the age of Tuc-Hor, then their measurement for stars unaffiliated with any moving group can distinguish analogs to stars in star-forming regions ($\tau = 1$--20 Myr) from the young ($\tau = 50$--300 Myr) field population \citep[][]{Shkolnik:2012ee} and old field stars \citep[][]{Reid:2004rw}.

The current census of Tuc-Hor is largely restricted to the higher-mass (AFGK-type) stars, which can be selected via all-sky activity indicators like ROSAT and confirmed with high-quality proper motions from Hipparcos. There are $<$10 spectroscopically confirmed M dwarfs in Tuc-Hor, even though these stars represent the peak of the IMF and thus should comprise the majority of the population by both number and mass. The reason for this paucity is straightforward. M dwarfs are fainter both optically and in the Xray/UV, so they have been more difficult to mine out of all-sky surveys. \citet[][]{Malo:2012fv} and \citet[][]{Rodriguez:2013vn} have begun to identify significant samples of low-mass candidate members, based on proper motions and ROSAT/GALEX excesses, but they spectroscopically confirmed only a handful of late-type members.

In this paper, we present the discovery and spectroscopic confirmation of 129 new K3--M6 members of the Tuc-Hor moving group, along with the recovery of most known $\ge$K3 members, and compute isochronal sequences for several spectroscopic signatures of youth. We also use this sample to characterize the age, mass function, spatial and velocity distribution, disk population, and activity of Tuc-Hor members. In Section 2, we describe our candidate selection procedures. In Section 3, we describe our high-resolution optical spectroscopic observations, and in Section 4, we summarize the analysis methods used to measure each candidate's spectroscopic properties. In Section 5, we combine all of the signatures of youth and membership to identify a sample of bona fide moving group members with spectral types mid-K to mid-M, and compare our results to those of previous surveys. Finally, in Section 6, we discuss the population statistics of the Tuc-Hor moving group.

\section{Candidate Selection}

 \begin{figure*}
 \epsscale{1.0}
 \plottwo{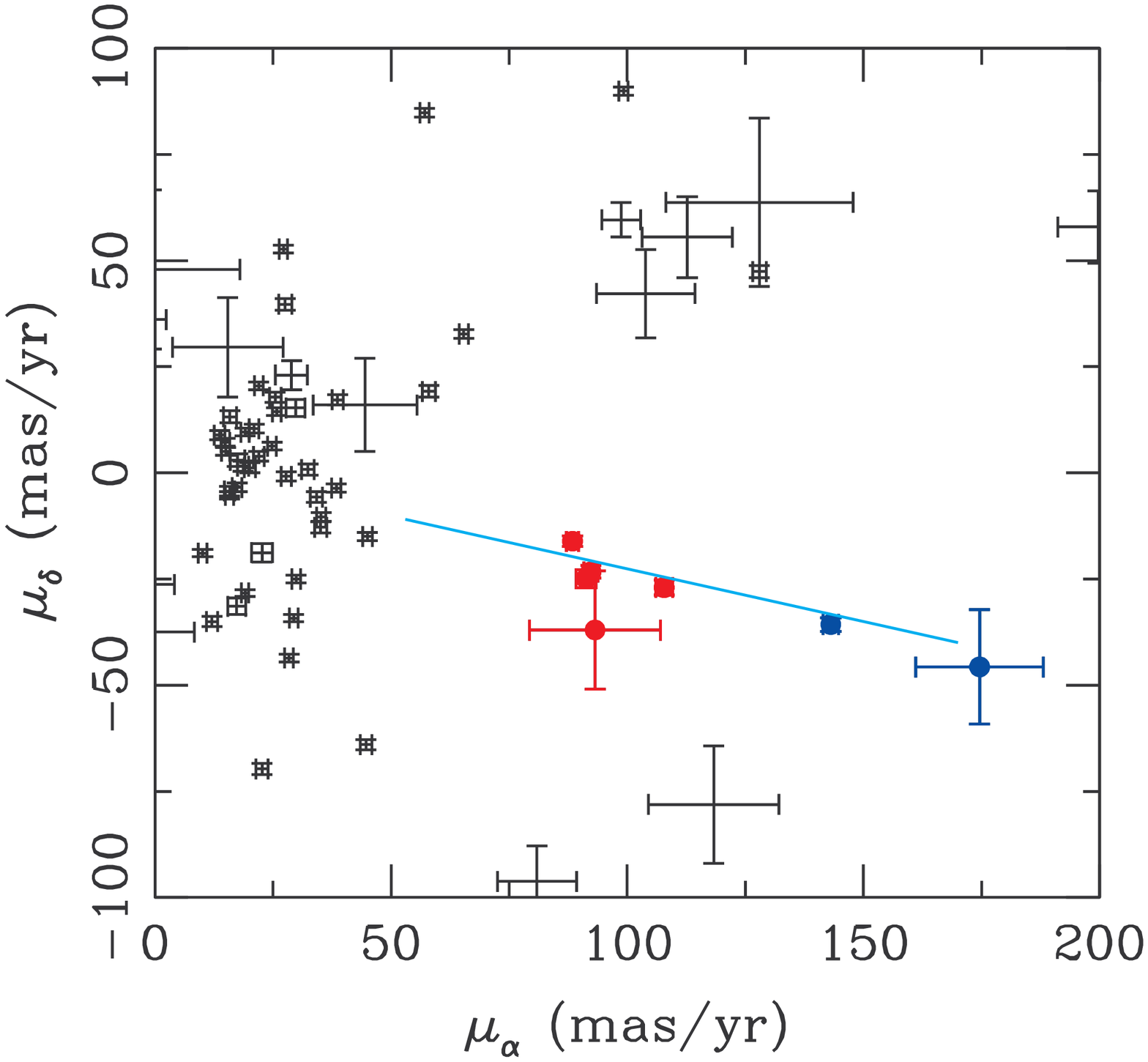}{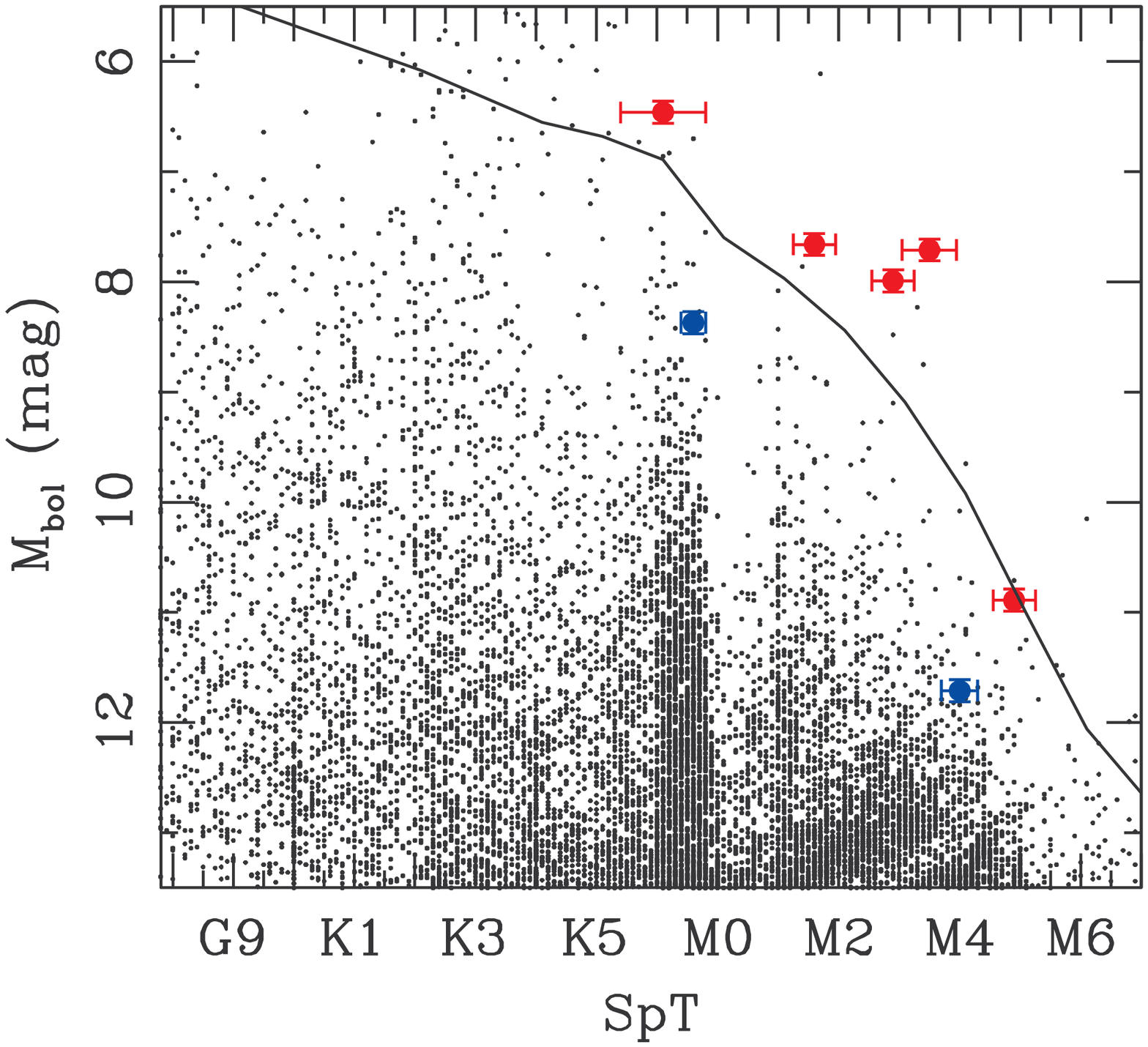}
 \caption{Illustration of our selection criteria for 16 deg$^2$ near the center of the Tuc-Hor locus on the sky ($26^o < \alpha < 34^o$ and $-60^o < \delta < -56^o$), showing that any bona fide moving group member must fall along the moving group locus in proper-motion space and above the main sequence in an HR diagram. Left: Proper motion diagram for all sources with spectrophotometric distance $d < 80$ pc. The expected locus for Tuc-Hor members along this line-of-sight with $20 < d < 80$ pc is denoted with a cyan line. Five candidates that pass our proper motion cut and fall above the MS at their purported kinematic distance are shown with red symbols, while two candidates that pass our proper motion cut and fall below the MS at their purported kinematic distance are shown with blue symbols. All remaining objects are shown with symbol-less error bars. Right: HR diagram for all objects we detected in this 16 deg$^2$ field, including the seven candidates show with colored points in the left panel. The absolute magnitude $M_{bol}$ is computed from the apparent bolometric magnitude ($m_{bol}$; Section 2.3) and the best-fit kinematic distance modulus ($DM_{kin}$; left panel and Section 2.4). The solid line shows the main sequence \cite[][]{Kraus:2007mz}. All five of the objects which pass our proper motion cut and fall above the MS for their kinematic distance were subsequently confirmed to be bona fide members, demonstrating the power of combined photometric-astrometric member searches. However, the clear structures within the background star population show that complicated SED-fit procedures must be interpreted with caution; we discuss this point further in Sections 2.3 and 6.1.}
 \end{figure*}

 \begin{figure*}
 \epsscale{1.0}
 \plotone{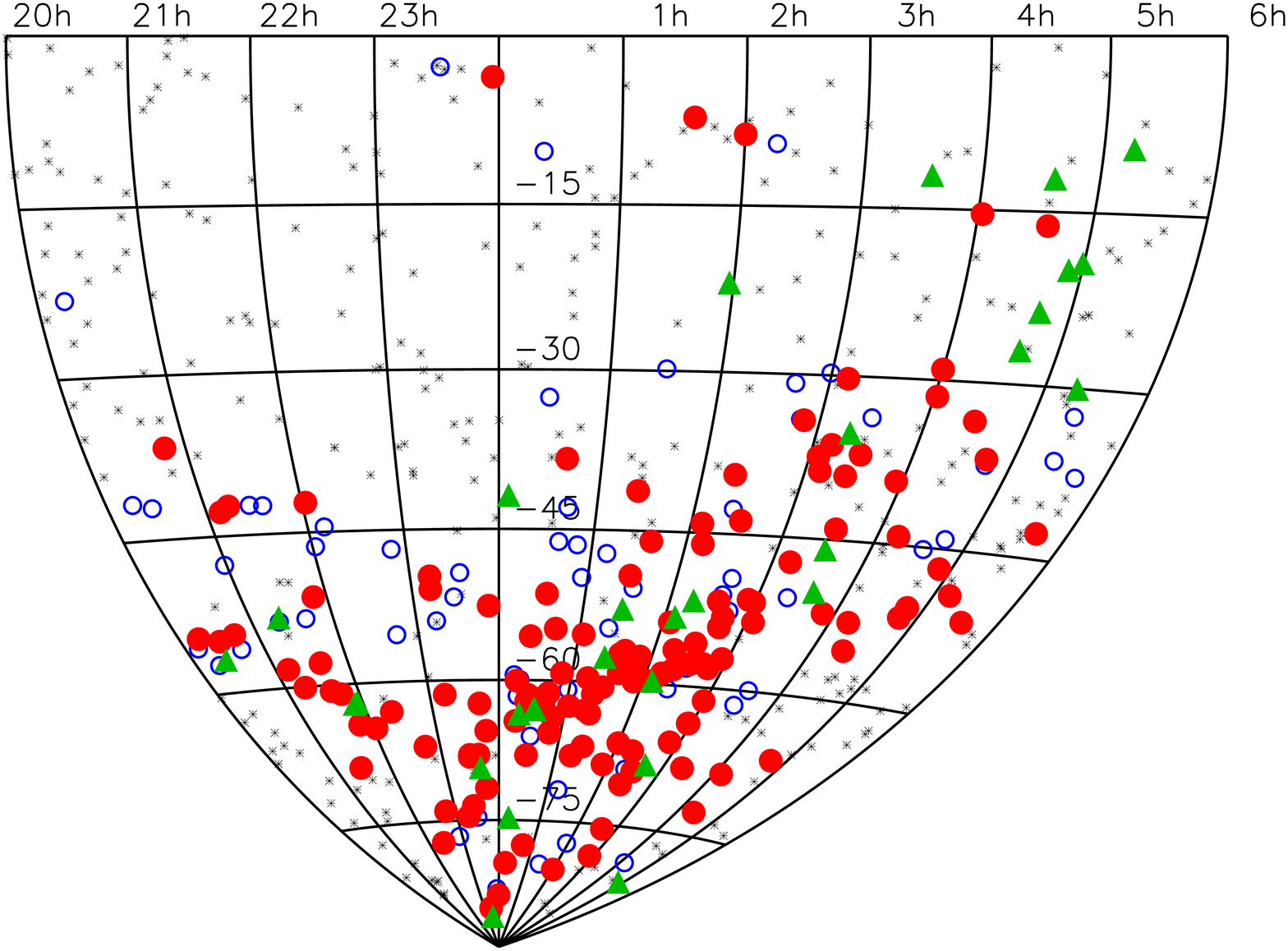}
 \caption{Positions of our candidate Tuc-Hor members on the sky, shown in Aitoff projection. Candidates which were confirmed to be bona fide members are shown with red filled circles, while disproven interloper field stars are shown with blue open circles. The remaining unobserved candidates are shown with small black asterisks. Known members with Spt$<$K3 \citep[][]{Torres:2008lr} are shown with filled green triangles. Most of the unobserved candidates are far from the locus of known Tuc-Hor members or are too faint to be efficiently observed with an optical echelle spectrograph ($R \ga 15$). We also observed a small number of northern candidates that are X-ray active, even though they were far from known members, in order to test if the member distribution extended northward.}
 \end{figure*}

Pre-main sequence stars in a stellar population can be distinguished by three features: common movement through space, overluminosity compared to the zero-age main sequence (ZAMS), and the presence of various spectroscopic, UV/X-ray, or mid-IR signatures of youth. For data mining of candidates across large swaths of the sky, the most cost-effective features to select against are the proper motion and the overluminosity, since both can be computed from existing all-sky survey data. As we describe below, we have combined multiple surveys in this work, using methods first described in \citet[][]{Kraus:2007mz}. Broadband optical-NIR photometry and proper motions also are largely unbiased against activity and disk existence, allowing for robust population statistics in the resulting member census. In particular, we did not use any activity criteria (such as Xray or UV emission) to select candidates because one of our primary goals was to test the efficiency and completeness of those selection methods (Sections 5.2 and 6.7).

In the following subsections, we list the all-sky surveys that contribute to our work, describe the calculation of proper motions, describe the calculation of bolometric fluxes and spectral types, and explain the selection of candidate members of Tuc-Hor. We selected our candidates from the entire southern sky ($\delta < 0^o$) between right ascensions of $20^h < \alpha < 6^h$, encompassing the entire spatial distribution of previously known members.

\subsection{Data Sources}

\subsubsection{USNO-B1.0}

The USNO-B1.0 survey (USNOB; \citealt[][]{Monet:2003yt}) is a catalog based on the digitization of photographic survey plates from five epochs. For fields north of $\delta = -30^o$, these plates are drawn from the two Palomar Observatory Sky Surveys, which observed the entire available sky in the 1950s with photographic B and R plates and the 1990s with photographic B, R, and I plates. For fields south of $\delta = -30^o$, including most of Tuc-Hor, the corresponding observations were taken by the UK Schmidt telescope in the 1970s-1980s and 1980s-1990s, respectively. 

The approximate detection limits of the USNOB catalog are $B \sim 20$, $R \sim 20$, and $I \sim 19$, and the observations saturate for stars brighter than $R \sim 11$. The typical astrometric accuracy at each epoch is $\sim$200 mas, dominated by systematic uncertainty due to its uncertain calibration into the International Celestial Reference System (ICRS) via the unpublished USNO YS4.0 catalog (as tested for specific pointings by \citealt[][]{Kraus:2007mz}, and verified across the entire sky by \citealt[][]{Roeser:2010fr}).

\subsubsection{2MASS}

The Two-Micron All-Sky Survey \citep[2MASS; ][]{Skrutskie:2006nb} observed the entire sky in the J, H, and Ks bands over the interval of 1998--2002. Each point on the sky was imaged six times and the coadded total integration time was 7.8 s, yielding 10$\sigma$ detection limits of $K_s = 14.3$, $H = 15.1$, and $J = 15.8$. The typical astrometric accuracy is $\sim$70 mas for any source detected at $S/N > 20$, as for all the sources considered in this work. The absolute astrometry calibration was calculated with respect to stars from Tycho-2; subsequent tests have shown that systematic errors are typically $<$30 mas \citep[][]{Zacharias:2004zi}.

\subsubsection{DENIS}

The Deep Near Infrared Survey of the Southern Sky \citep[DENIS; ][]{Epchtein:1994rt} observed $\sim$84\% of the southern sky (with some gaps) in the optical (with a Gunn $i$ filter) and the near-infrared (with $J$, and $K_s$ filters) during 1995-2001. It observed 3662 strips that each spanned 30$^o$ in declination and 12\arcmin\, in right ascension, reaching limiting magnitudes of $i=18.5$, $J=16.5$, and $K_s=14$ and saturating at $i=9.8$, $J=7.5$, and $K_s=6.0$. The photometry is nominally redundant with 2MASS (for $J$ and $K_s$), SDSS (for SDSS $i'$) and USNOB (for $I$, though the DENIS $i$ magnitude is more precise than the USNOB $I2$ magnitude), but is still useful for reducing stochastic errors and accounting for potential variability. Our experiments indicate that some strips from DENIS have significant systematic discrepancies in the tie-in to the ICRS, so we do not use DENIS astrometry in our proper motion calculations.

\subsubsection{SDSS}

The Sloan Digital Sky Survey \citep[SDSS; ][]{York:2000gd} is an ongoing deep optical imaging and spectroscopic survey of the northern galactic cap and selected regions of the southern cap. The most recent data release \citep[DR9; ][]{Ahn:2012ys} reported imaging results in five filters (ugriz) for 14,555 deg$^2$. The 10$\sigma$ detection limits in each filter are $u = 22.0$, $g = 22.2$, $r = 22.2$, $i = 21.3$, and $z = 20.5$; the saturation limit in all filters is $m \sim 14$. The typical absolute astrometric accuracy is $\sim$45 mas rms for sources brighter than $r \sim 20$, declining to 100 mas at $r \sim 22$ \citep[][]{Pier:2003pf}; absolute astrometry was calibrated with respect to stars from UCAC2, which is calibrated to the ICRS.

The default astrometry reported by the SDSS catalog is the r-band measurement, not the average of all five filters. However, the residuals for each filter (with respect to the default value) are available, so we used these residuals to construct a weighted mean value for our analysis. We adopted a conservative saturation limit of $m=15$ in all filters, even though the nominal saturation limit is $m=14$, because we found that many photometric measurements were mildly saturated for $14 < m < 14.5$. We also neglect measurements which are flagged by the SDSS database as having one or more saturated pixels. Finally, we removed all sources which did not have at least one measurement above the nominal 10$\sigma$ detection limits. Any moving group members fainter than this limit will not have counterparts in other catalogs, and the presence of excess sources can complicate attempts to match counterparts between data sets.

\subsubsection{UCAC3}

The astrometric quality of the above surveys could be compromised for bright, saturated stars, so proper motions calculated from those observations could be unreliable. Many of the brightest stars are saturated in all epochs, so we have no astrometry with which to compute proper motions. We have addressed this problem by adopting proper motions for bright stars as measured by the Third USNO CCD Astrograph Catalog \citep[UCAC3; ][]{Zacharias:2010lr}.

UCAC3 was compiled from a large number of photographic sky surveys and a complete reimaging of the sky by the U.S. Naval Observatory Twin Astrograph. UCAC3 extends to $R = 16$, though the proper motion errors become quite large at $R > 13$--14. The typical errors in the reported proper motions are $\sim$1--3 mas/yr down to $R=12$ and $\sim$6 mas/yr down to $R =16$. 

\subsection{Proper Motions}

Many recent efforts have employed various combinations of all-sky surveys in order to systematically measure proper motions of both clusters and field stars; USNOB is itself a product of such analysis, \citet[][]{Gould:2004wd} combined SDSS and USNOB, and the PPMXL survey \citep[][]{Roeser:2010fr} combined 2MASS and USNOB. However, there has been no systematic attempt to combine USNOB, 2MASS, and SDSS using a single algorithm to produce a single unified set of proper motions. All catalogs are calibrated into the ICRS, so in principle their coordinate lists can be adopted without any need for further calibration. In practice, this choice incurs a systematic error of $\sim$200 mas on each USNOB epoch (as described above), though 2MASS and SDSS appear highly consistent. 

We obtained the astrometry for all sources from the Vizier archive using the IDL routine queryvizier.pro, and then combined the coordinate lists for each source using a weighted least-squares fit. Our algorithm tested the goodness of each fit and rejected all outliers at $> 3 \sigma$; most of these outliers were found in the photographic survey data, not in 2MASS or SDSS, due to the heavy weight assigned to the modern CCD-based epochs. We find that the addition of at least one high-quality modern data point from 2MASS or SDSS reduces the uncertainties on a given proper motion by a factor of $\sim$2 compared to standard USNOB proper motions; the use of a sigma clip also substantially reduces the number of extremely erroneous measurements, since USNOB used no sigma clip. As we showed in \citet[][]{Kraus:2007mz}, this procedure led to a recovery rate of $>$90\% for the known members of Praesepe, and hence any population mined out of this dataset should be nearly complete.

Finally, we supplemented our measurements for bright stars with the proper motions from the UCAC3 catalog, which typically yields more precise proper motions for $R \la 12$ mag. In any case where the reported uncertainty from UCAC3 was less than the inferred uncertainty in our measurement, we simply adopted its measurement instead. A comparison of these measurements shows that they are consistent within the uncertainties. Combined with the $>$90\% yield we found for recovering open cluster members \citep[][]{Kraus:2007mz}, our proper motions appear broadly robust.

\subsection{Photometry and SED Fits}

Most population membership surveys select candidates based on agreement with the expected isochrone sequence in one or more color-magnitude diagrams. However, when many photometry points are available (i.e., 16 points for 2MASS+SDSS+DENIS+USNOB), then this procedure is unwieldy and neglects important covariances in the data. A superior method is to fit all available data with a model drawn from a grid of template SEDs, using all available photometry to find the best-fit stellar parameters ($T_{eff}$ and $f_{bol}$, or equivalently SpT and $m_{bol}$). If a star is assumed to fall on the main sequence, then a comparison of the inferred $m_{bol}$ with the expected $M_{bol}$ also directly yields a spectrophotometric distance modulus $DM_{phot}$. We describe the motivation and details in \citet[][]{Kraus:2007mz}. Our definition of the main sequence was tied to the Praesepe open cluster sequence, so the uncertainty in the corresponding value of $DM_{phot}$ is most likely dominated by the uncertainty in the distance modulus for Praesepe ($\sim$0.1 mag) and variations in the photometric calibration of USNOB for the northern and southern skies ($\sim$0.2--0.3 mag), since our USNOB-2MASS SEDs were bootstrapped from SDSS-2MASS SEDS for Praesepe. The corresponding uncertainty is therefore $\pm$0.2--0.3 mag in $M_{bol}$ or $\pm$0.2 subclasses in spectral type.

As for the astrometry, we obtained the photometry for all sources using queryvizier.pro and then computed the $\chi^2$ goodness of fit against a set of 541 main-sequence spectral type templates spanning B8.0-L5.0 in steps of 0.1 subclass. We describe the SED library and its construction in more detail in \citet[][]{Kraus:2007mz} and in Bowsher et al. (in prep).  We rejected potentially erroneous observations by identifying any measurement that disagreed with the best-fit SED by more than 3$\sigma$, where $\sigma$ is the photometric error reported by the CCD-based surveys or is adopted to be $\pm$0.25 mag for USNOB, and then calculating a new fit. The uncertainties in the spectral type and distance modulus were estimated from the $\Delta \chi^2 = 1$ interval of the $\chi^2$ fit for each object. 

The distribution of reduced $\chi^2$ values for our fits had too many sources with $\chi_{\nu}^2<1$ and $\chi_{\nu}^2>1$, and too few with $\chi_{\nu}^2 \sim 1$. Further investigation revealed the source of this discrepancy to be apparent non-Gaussianity of the errors of USNOB photometry. While the ``typical'' uncertainty is indeed $\pm$0.25 mag, this uncertainty actually consists of a stochastic component with $\sigma < 0.25$ mag and a systematic uncertainty (most likely due to the photometric calibration for individual photographic plates) which can exceed 1 magnitude. A global recalibration of USNOB would likely reduce this systematic uncertainty, but is beyond the scope of the current work.

As we discuss in Section 4.1, a comparison of spectral types from SED fits and from spectroscopy yields excellent agreement, with dispersion and systematic offsets of $\la$1 subclass for bona fide members across the K-M spectral type range. Given a typical uncertainty of at least 0.5 subclass for most spectroscopic spectral types (including those of our standard stars), then the observed dispersion in the relation indicates that photometric SpTs can be measured with similar accuracy. 

However, the HR diagram that we show in Figure 1 suggests that systematic errors remain for some spectral type ranges. In particular, it appears that M0-M1 stars might be systematically pulled to a classification of $\sim$K7.5, perhaps due to an error in the SED grid. This error was not seen in northern populations (e.g., \citealt[][]{Kraus:2007mz}), but those fields also had SDSS data that dominated the fits. Without SDSS data, then USNOB and 2MASS colors become more significant in the fits. We are producing updated SED templates for use in future surveys, but since we selected our input sample for this study with the old templates, then we have retained them for the present analysis. As we discuss further in Section 6.1, this systematic error could account for the absence of M0-M2 members within our sample.

\subsection{Selection Criteria}

After computing proper motions and SED fits, our input set consisted of $1.2 \times 10^8$ sources spanning 8300 deg$^2$ to consider as potential cluster members. However, almost all have proper motions inconsistent with membership or are too faint and blue to sit on the moving group sequence, so this number can be efficiently winnowed down. Unlike for compact clusters, moving group members span a large range of distances and a large area of the sky, and hence do not share a single proper motion. The kinematic selection must instead be made against the projection of the moving group's $UVW$ space velocity onto the plane of the sky at each source's position. Furthermore, the unknown distance means that the magnitude of the proper motion is a free parameter. Our selection of candidates from this input set can be divided into two main stages.

First, for each source we computed the angle of the expected proper motion, given the $UVW$ space motion of Tuc-Hor ($U = -9.9 \pm 1.5$ km/s, $V = -20.9 \pm 0.8$ km/s; $W = -1.4 \pm 0.9$ km/s; \citealt[][]{Torres:2008lr}). We then found the magnitude of the proper motion (i.e., the assumed distance) which minimizes the difference between the observed proper motion and the expected proper motion. We were then left with two quantities for each star: the discrepancy $\Delta$ (in mas/yr) between the observed proper motion and the best-fit proper motion it needed to have if it were a member, and the corresponding best-fit kinematic distance modulus $DM_{kin}$ corresponding to the magnitude of that best-fit proper motion vector. This method is similar to that used by \citet[][]{Lepine:2009pb} and \citet[][]{Schlieder:2010la}. We identified 1813 sources as kinematic candidates where the observed and expected proper motions agreed within 3$\sigma$ ($\Delta/ \sigma_{\mu} < 3$, where $\sigma_{\mu}$ is the observational uncertainty in the proper motion) or the total magnitude of the discrepancy was $\Delta < 10$ mas/yr, as well as requiring the kinematic distance to be $d < 80$ pc ($DM_{kin} \le 4.5$).

Second, for these 1813 kinematic candidates we computed the difference between the kinematic distance modulus $DM_{kin}$ and the spectrophotometric distance modulus $DM_{phot}$ to test whether, if an object were a member with the appropriate $DM_{kin}$, then it would sit at an appropriate height above the ZAMS. For our initial reconnaissance of this sample, we used the known members of Tuc-Hor \citep[][]{Torres:2008lr} to set the criterion for rejection at $DM_{kin}-DM_{phot} \le 0.0$. We also rejected any star with $DM_{kin}-DM_{phot} \ge 4.0$ since it would be too high above the ZAMS to be a member, and hence most likely is a field interloper with spurious $DM_{kin}$ and/or a giant with spurious $DM_{phot}$. We found that most candidates which were ultimately confirmed sat $\sim$0.5--1.0 mag above the main sequence, suggesting that our cuts should only miss a small number of candidates that sit low in the HR diagram. These cuts yielded 768 photometric/kinematic candidates. Finally, to produce a manageable sample for our observing time, we omitted any sources earlier than K3 (which should have been identified via HIPPARCOS in previous searches, and are heavily contaminated by subgiants) or later than M6 (which are too faint for efficient optical spectroscopy), leaving a total sample of 497 potential low-mass Tuc-Hor members.

In Figure 1, we illustrate the astrometric and photometric selection procedures for an area of 16 deg$^2$ near the central locus of the known and new Tuc-Hor nenvers. In Figure 2, we show a map of our candidates on the sky, including those which we observed and ultimately found to be either bona fide Tuc-Hor members or field star interlopers. Within the full 8300 deg$^2$ area, we prioritized those stars which were closest to the known members and which were bright enough to be observed with $\le$15 minute integration times. Our survey did not clearly reach a boundary for the distribution of Tuc-Hor members on the sky, and indeed, it is unclear whether even our initial area of 8300 deg$^2$ is sufficient to encompass all Tuc-Hor members, so future reconnaissance of a wider area might be necessary. Given the flux limits of the input all-sky surveys ($R \la 18$ mag, $K_s \la 14$ mag), then our input sample should have included $\tau = 30$ Myr Tuc-Hor members down to the substellar boundary ($M = 0.07 M_{\odot}$) at a distance of $d = 80$ pc. The flux limit for our optical spectroscopy ($R \sim 15$ mag) raises this limit to $M = 0.15 M_{\odot}$ (SpT$\sim$M5) at $d = 80$ pc.

The SACY search of Tuc-Hor \citep[][]{Torres:2006fk,Torres:2008lr} found 17 members with spectral types of K3-M0, of which we recovered 15 as candidates. Our search missed HD 222259 B because it is a close companion to the G6V star DS Tuc, which affected its measurements in the input all-sky catalogs. CD-35 1167 would have been selected by our photometric selection procedures, but it has no UCAC3 counterpart and was too bright for our procedure to calculate a new proper motion, so our astrometric selection procedure missed it.  We therefore estimate that our candidate list is $88^{+4}_{-12}\%$ complete in this spectral type range. There are few known members of Tuc-Hor later than M0, so we can not estimate the completeness of our sample, though we discuss the recovery of these members in Section 5.2. However, our results for Praesepe and Coma Ber remained $\ge$90\% complete to nearly the flux limit of USNO-B1 ($R\sim19$ mag), so we do not expect the completeness to differ substantially to the limit of our current Tuc-Hor sample ($R \sim 15$ mag). 

We identified CD-46 1064, BD-19 1062, and CD-30 2310 as candidates, but did not obtain spectra for them, so for uniformity we will not include them in the analysis summarized in Section 6. We obtained a spectrum of CD-35 1167 (which we missed as a candidate), and so we report its measurements and its confirmation as a member, but also will not include it in our analysis. We also obtained spectra for the other 12 previously known members in this list and will include them in our analysis.

Finally, we note that the Columba association is nearly comoving with Tuc-Hor in $UVW$ space and is coincident with the eastern end of Tuc-Hor on the sky, and hence we might expect some confusion between the two populations. However, the radial velocities differ by $\sim$4 km/s near the center of the point of overlap ($\alpha = 60^o$, $\delta = -45^o$), which can be distinguished at 3$\sigma$ for most of our candidates. Furthermore, the populations are distinct in XYZ space, with most known Columba members falling at $d > 60$ pc (\citealt[][]{Zuckerman:2004fj,Torres:2008lr}. There is no clear evidence of a parallel population meeting the RV or XYZ values of Columba (Sections 5.1 and 6.4), though this possibility should be considered after future surveys have more robustly determined the spatial and kinematic distributions off Columba. The handful of young stars that we find with $Z < -50$, which fall closer to the Columba locus, could be preliminary evidence of this overlap.

\section{Observations and Data Reduction}

Our candidates were selected by their proper motions and photometry, so we require independent measurements in order to confirm their membership in Tuc-Hor. The traditional methods for confirmation of young stars in stellar populations are to measure their radial velocity (testing for comovement in the dimension perpendicular to proper motion) and the identification of spectral signatures that can be associated with youth (such as lithium absorption, low-gravity diagnostics like shallow alkali lines, H$\alpha$ emission, and rapid rotation). In both cases, we can obtain the necessary measurements from high-resolution optical spectroscopy. Many of the diagnostics of youth have only been firmly calibrated for higher-mass stars \citep[e.g.,][]{Mentuch:2008vn}, so our measurements also yield the first robust calibration of isochronal parameter sequences for mid-K to mid-M dwarfs in Tuc-Hor.

We obtained high-resolution spectra for our targets in three observing runs in 2012 July, 2012 September, and 2013 February. We observed our targets using the Magellan Inamori Kyocera Echelle (MIKE) optical echelle spectrograph on the Clay telescope at Magellan Observatory. For all observations we used the 0.7\arcsec\, slit, which yields spectral resolution of $R=35,000$ across a range of $\lambda = 3350$-9500\AA. Since our targets are relatively red, most only had useable signal on the red chip ($\lambda > 5000$ \AA). The pixel scale oversamples the resolution with the 0.7\arcsec\, slit, so we observed with 2x binning in the spatial and spectral directions to reduce readout overheads. We also observed standard stars nightly from the list of \citet[][]{Nidever:2002gl}, which serve as both RV and spectral type templates, as well as most of the known Tuc-Hor members with SpT$\ge$K3, numerous members of the Sco-Cen OB association, and a selection of known members from other nearby young moving groups.

We reduced the raw spectra using the CarPy pipeline (Kelson 2003)\footnote{\url{http://code.obs.carnegiescience.edu/mike}} but used observations of spectrophotometric standard stars to measure and flatten the blaze function due to the uncertain temperature of the flat lamp. In order to correct for residual wavelength errors (due to flexure and uneven slit illumination), we then cross-correlated the 7600 \AA\, telluric A band for each spectrum against a well-exposed spectrum of a telluric standard, solving for the shift (typically $\sim$1 km/s) that places each spectrum into a common wavelength system defined by the atmosphere. Finally, we calculated and applied heliocentric radial velocity corrections. Multiple observations of dwarf standards suggest a repeatability of $<$0.5 km/s for our observations, as do multi-night observations of the young star 1SWASP J140747.93-394542.6 that will be reported in a future publication (M. Kenworthy et al., in preparation).

In Table 1, we list the epochs and exposure times for all of our MIKE observations of known or candidate Tuc-Hor members, as well as the $S/N$ for each spectrum at 6600 \AA. We also list all other relevant measurements used in target selection: proper motion, SED-fit SpT and bolometric flux, photometric and kinematic distance modulus, and the proper motion residual $\Delta$.

\section{Data Analysis}

 \begin{figure*}
 \epsscale{1.0}
 \plottwo{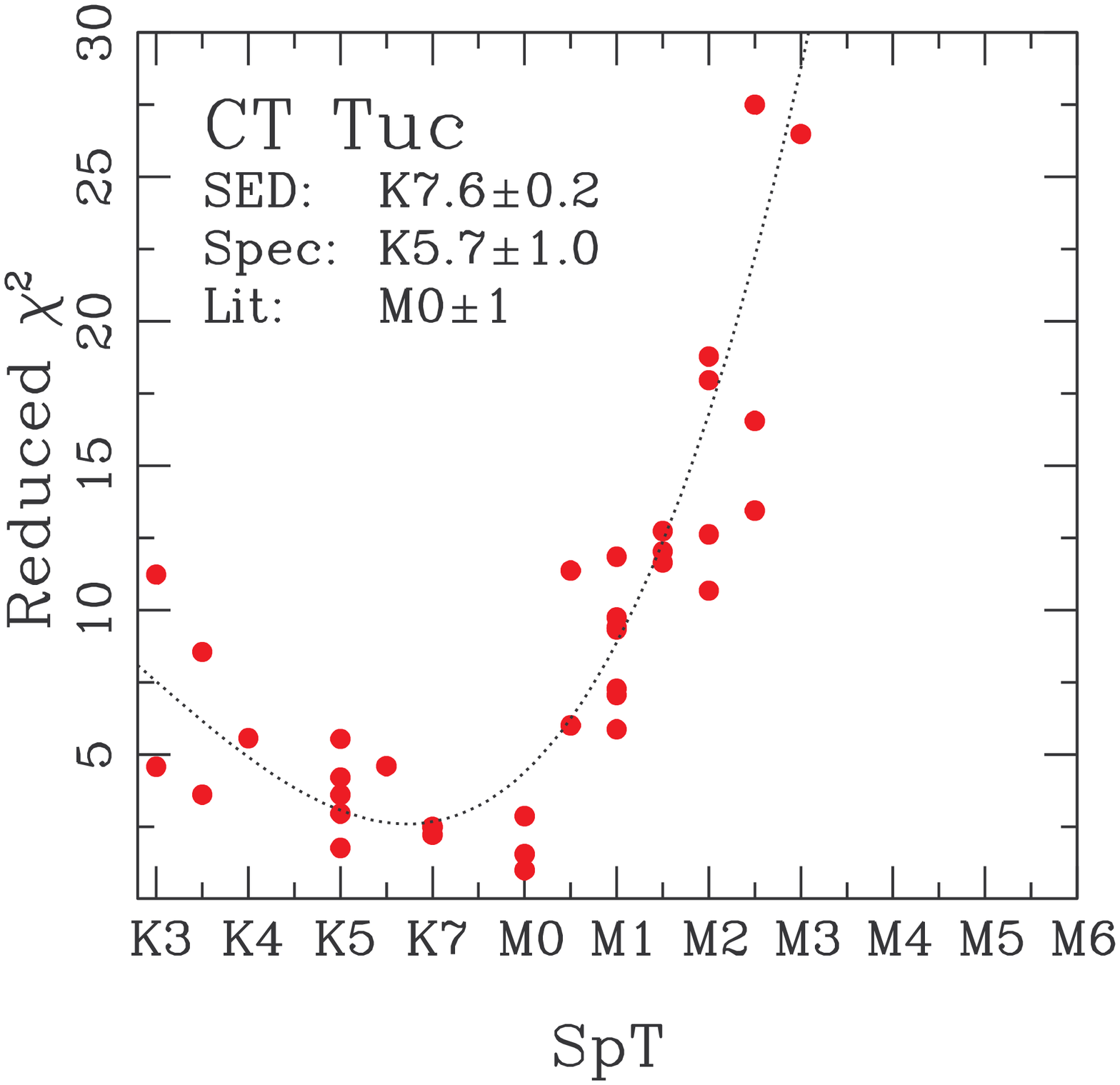}{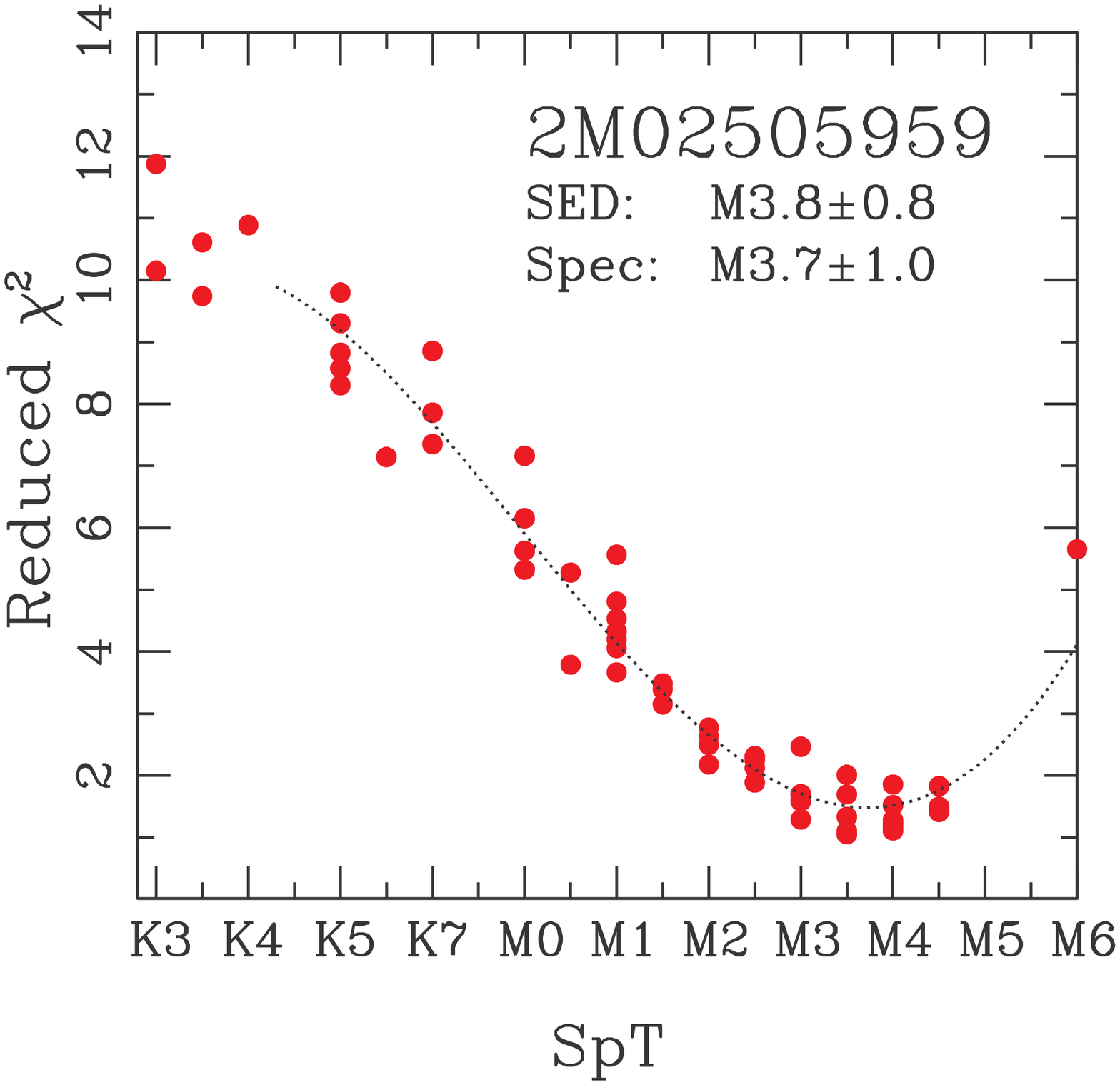}
 \caption{Reduced $\chi^2$ as a function of spectral type for comparisons between two target stars (CT Tuc, left, and 2MASS J02505959-3409050, right) and our grid of SpT standards. In each case, the dotted line shows the best-fit polynomial used to measure the spectral type. We also label the best-fit SpTs from our SED fits and our MIKE spectra, and in the case of CT Tuc, the previously assessed value from the literature (Zuckerman \& Song 2004).}
 \end{figure*}
 
  \begin{figure*}
 \epsscale{0.90}
 \plotone{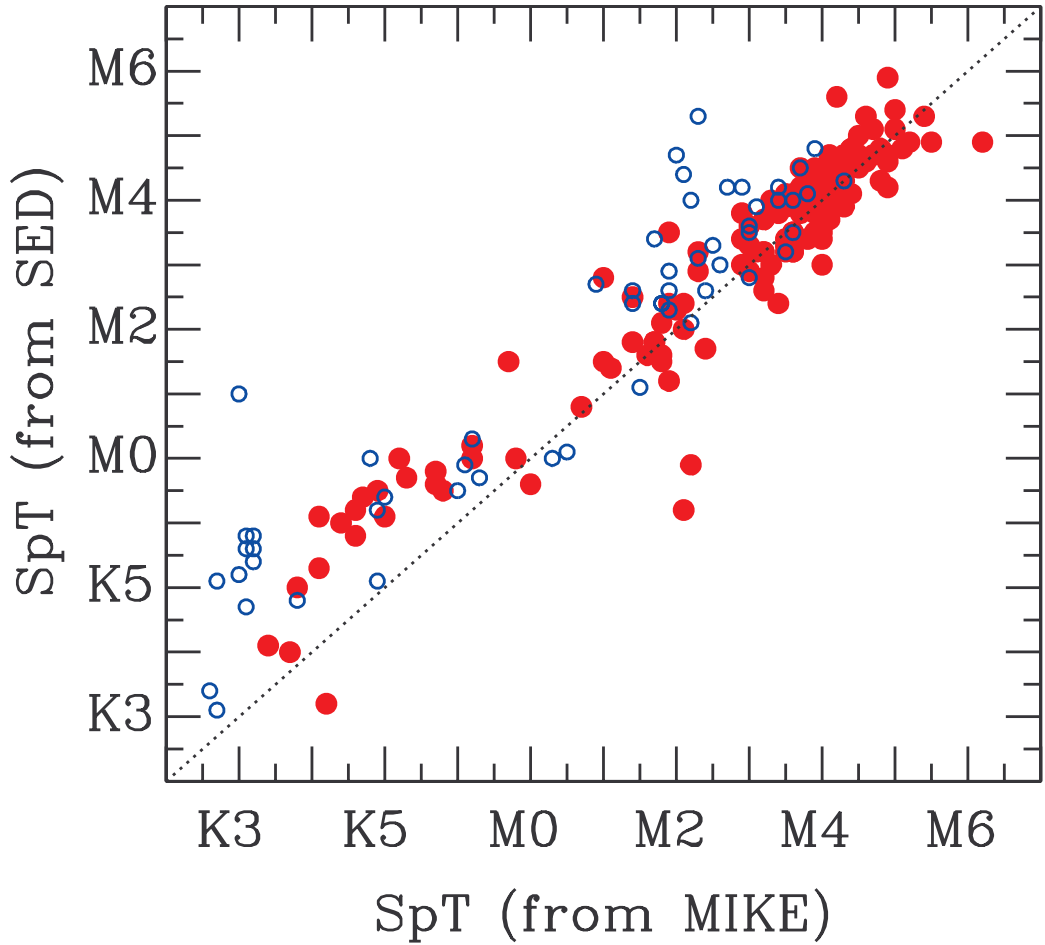}
 \caption{Comparison of spectral types estimated via our SED fit (Section 2.3) and via our spectroscopic observations (Section 4.1). Objects that were ultimately confirmed as members are shown with filled red circles, while objects assessed to be field interlopers are shown as open blue circles. For the 144 members with best-fit spectral types of $>$K3.0, the assessments agree with an offset of 0.2 subclass and a dispersion of 0.7 subclass. The nonmembers with SED SpTs of K5--K7 and spectroscopic SpTs of $\sim$K3 are background giants. Our SED-spectral types might be biased later by mild reddening from interstellar dust, while our spectroscopic spectral types (which are derived from line shapes in continuum-flattened spectra) are independent of reddening. The offset of +1 subclass between SED and spectroscopic SpTs for K4-K7 stars could be a result of differences in classification systems, which traditionally have irregularities at the K/M boundary depending on the definition and usage of K6--K9 spectral types (e.g., \citealt[][]{Schmidt-Kaler:1992ab} versus \citealt[][]{Pecaut:2013zr}), though the horizontal clustering seen in our HR diagram (Figure 1) suggests that our SED fits also should be examined more closely.}
 \end{figure*}
 
  \begin{figure*}
 \epsscale{1.0}
 \plotone{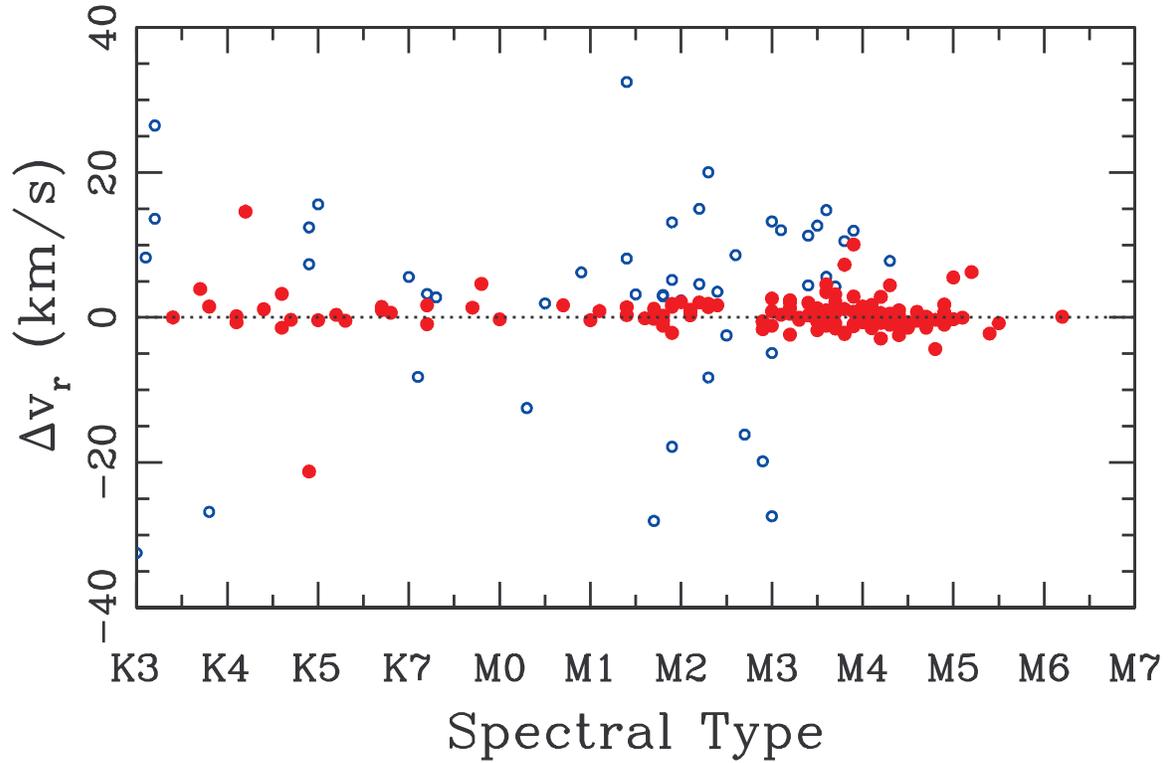}
 \caption{The RV residual ($\Delta v_{rad} = v_{rad,obs}-v_{rad,TH}$) for our observed targets as a function of (spectroscopically determined) spectral type. As in Figure 4, members are denoted with filled red circles and nonmembers with open blue circles. The substantial overdensity at $\Delta v_{rad} = 0$ indicates the bona fide members of Tuc-Hor. Some members should have discrepant RVs due to binarity; two of these SBs among the K dwarfs can be identified as Tuc-Hor members by the presence of $Li_{6708}$ absorption.}
 \end{figure*}
 
  \begin{figure*}
 \epsscale{1.0}
 \plotone{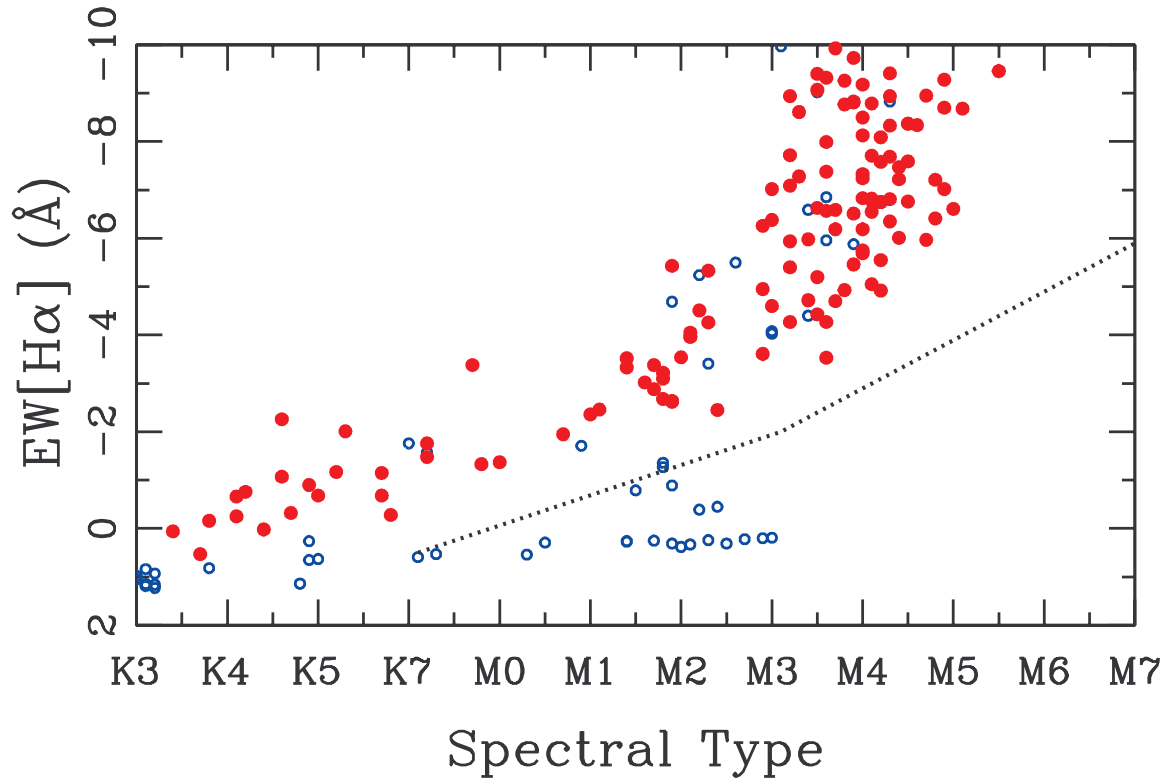}
 \caption{The H$\alpha$ equivalent width as a function of (spectroscopically determined) spectral type for our observed targets. As in Figure 4, members are denoted with filled red circles and nonmembers with open blue circles. The dotted line shows the lower envelope for H$\alpha$ emission from members of the contemporaneous open clusters IC 2602 and IC 2391 \citep[][]{Stauffer:1997ly}. Many nonmembers fall within the H$\alpha$ locus for Tuc-Hor, which suggests that they might be single-line spectroscopic binaries that fail our RV selection due to orbital motion.}
  \end{figure*}

 \begin{figure*}
 \epsscale{1.0}
 \plotone{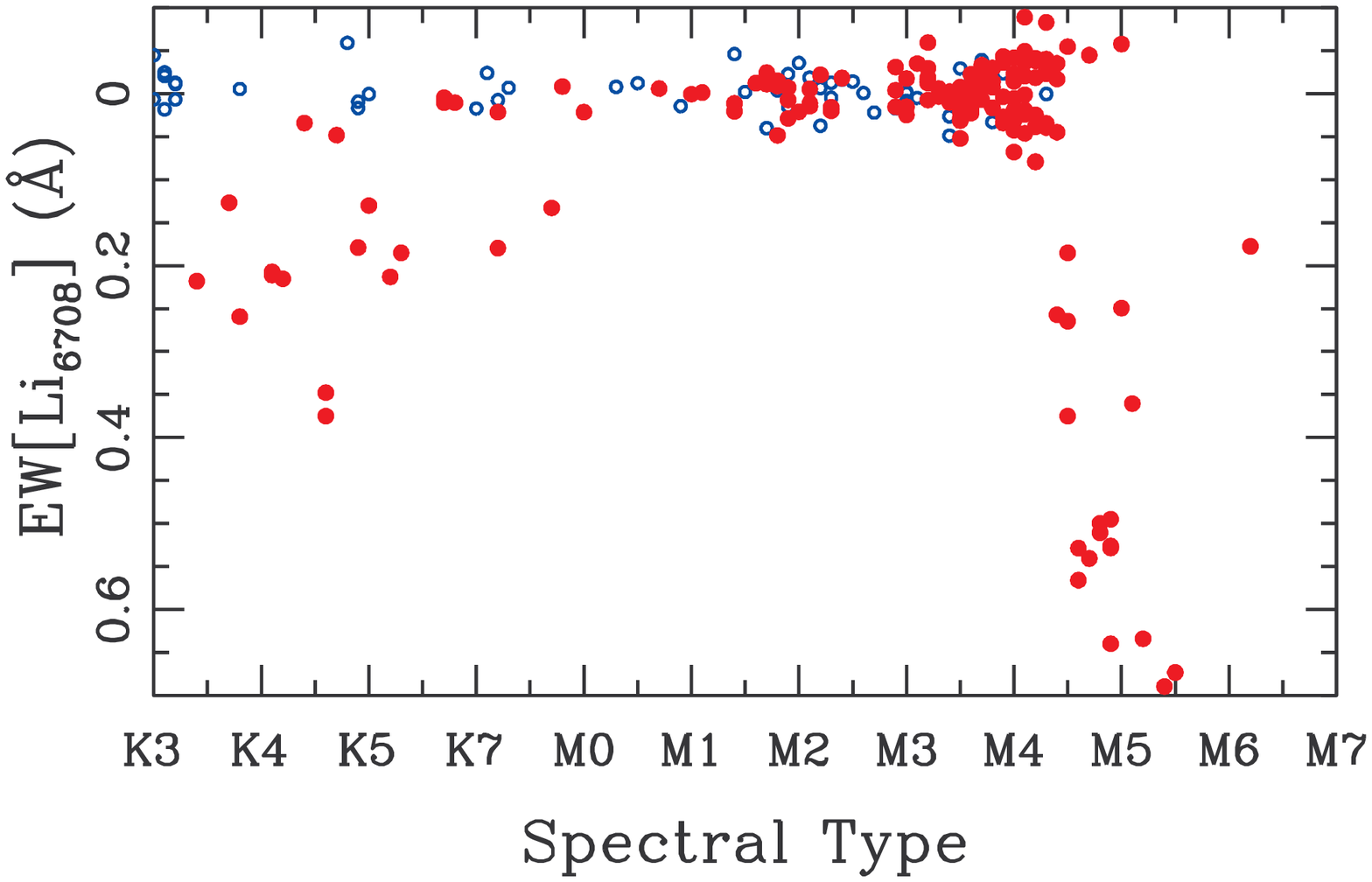}
 \caption{The Li$_{6708}$ equivalent width as a function of (spectroscopically determined) spectral type for our observed targets. As in Figure 4, members are denoted with filled red circles and nonmembers with open blue circles.}
  \end{figure*}

\subsection{Spectral Types}

Our spectra provide a valuable check of the SpTs estimated from SED fitting (Section 2.3). Our SED SpTs broadly match known spectroscopic SpTs with a dispersion of 0.7 subclasses, but erroneous input photometry could significantly bias the results for some individual stars. To measure spectral types for our targets, we computed the $\chi^2$ goodness of fit with respect to standard stars of known spectral type. We applied this analysis using both primary standards (well-studied dwarfs with known RV and SpT, drawn from the sample of \citealt[][]{Nidever:2002gl}) and secondary standards (67 candidate Tuc-Hor members for which spectroscopic SpTs are available in the literature). The absolute value of the $\chi^2$ statistic can not be easily interpreted because each pixel of spectrum conveys a different amount of information about a star (with much distinct information from temperature-sensitive lines or molecular bands, and little distinct information from temperature-independent lines or from continuum). However, the relative $\chi^2$ values can be used to determine which standard spectrum best matches a given science target.

Each science target spectrum was compared to all standard star spectra. In this procedure, the standard spectra were shifted to the same radial velocity (using RV measurements from Section 4.2), the spectrum with narrower lines was convolved with a Gaussian kernel to match the more broadened spectrum (based on $v_{rot}$ measurements from Section 4.2), and then the reduced $\chi^2$ values were computed for each of seven orders ($\lambda = 6100$\AA, 6700\AA, 7000\AA, 7100\AA, 7400\AA, 7900\AA, and 8400\AA). We adjusted the assumed uncertainties for each order so that the best-fitting standard yielded reduced $\chi^2 = 1$, recomputed all fits, and then averaged the reduced $\chi^2$ values across all seven orders to find the standard star that best fit the entire spectrum. Finally, since the standard stars are quantized by 0.5 or 1.0 subclasses, we fit the set of reduced $\chi^2$ values (as a function of standard-star SpT) with a low-order polynomial in order to find the true minimum in the relation (indicating the ideal best-fit SpT).

In Figure 3, we demonstrate the results of this fit for two stars in our sample. The known Tuc-Hor member CT Tuc was assessed to be an M0V star by Zuckerman \& Song (2004), and our SED fit yielded a SpT of K7.6. Our spectroscopic analysis finds that the best-fit standard is an M0 star (HIP 1910, another Tuc-Hor member), while a polynomial fit of the reduced $\chi^2$ surface yields a best-fit SpT of K5.7. Similarly, 2MASS J02505959-3409050 (a previously unidentified candidate member) is found to be an M3.8 star in our SED fit, whereas our spectral analysis finds it is best fit by an M3.5 standard star and the polynomial fit of the reduced $\chi^2$ surface yields a best-fit SpT of M3.7. In Figure 4, we compare our spectroscopic SpTs with the SED SpTs computed in Section 2.3, and show that the two results broadly agree for $\ge$M0 dwarfs. We find that stars which appear spectroscopically to be K3-K7 dwarfs are fit with SED SpTs which are systematically $\sim$1.5 subclasses later. Mid-late K dwarfs are defined in various ambiguous ways (with variable usage of the K6/K8/K9 types) and span a wide range of $T_{eff}$, so this systematic uncertainty is not unexpected; until more independent spectroscopic studies of these stars are available, we can not determine whether this systematic offset results from the color-SpT relations that we used to compute SED fits or from our choice of spectroscopic standards. We also found that one Tuc-Hor member (2MASS J20474501-3635409) has a best-fit spectral type of G7.4, substantially earlier than the SED-fit spectral type of K4.3. This star is a very rapid rotator ($\sigma_{vsin(i)} = 106$ km/s), so a spectral classification from spectral line strengths is highly uncertain; the SED-fit spectral type is likely to be more valid, so we retain it in our K3--M6 sample.

In Table 2, we report the best-fitting polynomial-fit SpTs for each candidate Tuc-Hor member.

\subsection{RVs, Rotational Broadening, and SB2s}

We measured radial velocities for our targets using broadening function deconvolution \citep[BFD; ][]{Rucinski:1999yr}; as Rucinski described, broadening functions have a flatter base than cross-correlations and are less susceptible to effects like ``peak pulling'' for spectroscopic binaries. S. Rucinski distributes an IDL pipeline that is designed to conduct BF for any input spectrum, and we adopted this pipeline as written.\footnote{\url{http://www.astro.utoronto.ca/~rucinski/}}

For each order of each spectrum, we used the BFD pipeline to calculate the broadening function with respect to a bright RV standard star that best matches the target star's spectral type (Sections 2.3 and 5.1). We then fit the peak of the broadening function with a Gaussian function in order to measure the RV in that order, and measured the average RV for that spectrum by calculating a mean of all orders with $S/N > 7$ at the order midpoint. We observed some dwarf templates and young stars multiple times, and we found that the typical standard deviation across multiple epochs was $\sigma \sim 0.4$ km/s, indicating that any systematic noise floor falls at or below this limit. Finally, we computed the rotational or instrumental broadening (the standard deviation $\sigma$ of the Gaussian fits) by calculating the mean value of $\sigma$ for all Gaussian fits across the same orders. The spectral resolution of $R=35000$ corresponds to a minimum detectable broadening of $\sigma \sim$4.5 km/s.

Finally, we found six SB2s and three SB3s among the 206 candidates that we observed. We fit these targets with a two- and three-component versions of our BFD pipeline, and determined separate radial and rotational velocities for each component. We also fit the ratio of the cross-correlation areas (a proxy for flux ratio) as a function of wavelength and used a linear fit to estimate the flux ratio at 7600\AA\, (the SDSS $i'$ filter). This process is described in more detail in \citet[][]{Kraus:2011lr} and \citet[][]{Law:2012qa}.

For the SB3s, we assumed that blueshifted and redshifted components were in a close pair, while the intermediate-velocity component was likely the singleton tertiary; the velocity of this tertiary was then adopted as the best available estimate of the system velocity. For the SB2s, we used the flux ratio to assess an approximate mass ratio (and hence ratio of RV amplitudes) using the 30 Myr models of \citet[][]{Baraffe:1998yo}, and then combined the ratio of RV amplitudes with the total RV difference in order to estimate the system velocity. These system velocities are more uncertain due to the significant systematic uncertainties in pre-main sequence evolutionary models (e.g., Hillenbrand \& White 2004), but based on the members which can be independently confirmed from lithium or H$\alpha$ (and hence should be comoving with Tuc-Hor), we estimate that they should be reliable to within $\pm$2 km/s.

We list the stellar or system velocities, the discrepancy ($\Delta v_{rad}$) compared to the expected values for Tuc-Hor members, and the rotational broadening values in Table 2, and report the properties of all SB2s in Table 3 and SB3s in Table 4. In Figure 5, we plot the difference $\Delta v_{rad}$ between the observed RV and the expected RV (for a Tuc-Hor member at that position on the sky) as a function of spectral type; there is a clear excess at $\Delta v = 0$ km/s, denoting the Tuc-Hor population.  Some outliers can be identified as young stars via age diagnostics and hence are likely SB1s. Three apparently comoving stars appear old according to those same age diagnostics, and hence are likely field interlopers that happen to be comoving with Tuc-Hor.

\subsection{H$\alpha$ and Lithium}

Most young stars with SpT $\ge$M0 show the Balmer series in emission, and even earlier-type stars often show filled-in absorption lines (due to activity) or occasionally very broad emission (if they are still accreting from a disk). Due to the lack of other significant lines in this region of the spectrum, measurement of $EW[H\alpha]$ is quite straightforward for the vast majority of cases; we simply fit the absorption or emission line with a single Gaussian. In spectra where the morphology was complicated or $EW[H\alpha] \sim 0$, we instead measured the EW by setting the continuum level using sidebands spanning 2--5 \AA\, on either side, then summing the flux across the H$\alpha$ line to directly determine the excess or deficit. We also used this procedure to measure the total H$\alpha$ emission for all SB2s and SB3s, as the lines were generally too blended to confidently disentangle.

Another key indicator of youth is the lithium line at 6708 \AA, as lithium is rapidly burned at the base of the convective envelopes of late-K and M dwarfs as they age. At an age of $\tau \sim 30$ Myr, lithium should be completely depleted for stars of SpT K7--M4, but not yet burned for earlier and later types \citep[][]{Randich:2001lr,Dobbie:2010oq}. The measurement of $EW[Li_{6708}]$ is more complicated than for H$\alpha$, because it is blended with a weak Fe I line for G-K stars and has several other features in close proximity; for M dwarfs, the spectrum is quite complex. We therefore measured $EW[Li]$ by fitting that order of each spectrum with the dwarf template most closely resembling it (Section 4.1), then measuring the relative flux deficit for the science target within $\pm$1.0\AA\, of the expected wavelength. Only one SB (2MASS J05332558-5117131) showed lithium absorption in its spectrum, and it was only detectable at the expected wavelength for the primary star, so there was no need to conduct multiple-line fits to determine individual component line strengths. For 4 rapid rotators, we measured the equivalent widths manually using the IRAF task splot.

We found that when K stars otherwise appeared old (with discrepant RVs and no H$\alpha$ emission), then the mean equivalent width was $EW[Li] = 3$ m\AA, with a standard deviation of 16 m\AA. Any measurement with $EW[Li] > 50$ mA is significant at a confidence level of 3$\sigma$, and hence can be regarded as a confident detection. For M1--M3 stars (which we do not expect to host lithium at this age), the mean equivalent width was $EW[Li] = -1$ m\AA, with a standard deviation of 23 m\AA. The corresponding 3$\sigma$ limit is therefore $EW[Li] > 70$ m\AA.

We report our measurements for H$\alpha$ and $Li_{6708}$ in Table 2, and plot the equivalent widths as a function of (SED-fit) spectral type in Figures 6 and 7.

\section{The Membership of Tucana-Horologium}

\subsection{New Members}

Synthesizing our observations into a unified membership census is a complicated task, because any one measurement could yield a false positive or negative. Some field stars will be comoving in radial velocity by chance, and most short-period binaries among the bona fide Tuc-Hor members will not appear comoving in any single-epoch spectrum. Also, activity signatures show wide variations in any single-aged population \citep[][Shkolnik \& Barman, in preparation]{Stelzer:2012vn}, though some show a lower bound that allows for dispositive rejection of nonmembers. The most conclusively affirmative or dispositive measurement is the presence or absence of lithium, but this is only valid for a restricted range of spectral types. We use three criteria to select members of Tuc-Hor and reject likely field interlopers. In order of precedence, these criteria are:

\begin{itemize}

\item {\em Lithium:} For the assumed age of Tuc-Hor ($\tau \sim 20$--50 Myr), lithium should be depleted in the atmospheres of stars with spectral types $\ga$K7 and $\la$M5 \citep[][]{Mentuch:2008vn}. However, lithium is depleted across the entire spectral type range of our sample (K3--M6) by the age of $\sim$125 Myr (as seen in the Pleiades and AB Dor; \citealt[][]{Stauffer:1998mz,Mentuch:2008vn}). We therefore use the presence of lithium (with $EW[Li] > 100$ m\AA) as a youth indicator across our entire spectral type range (confirming 34 members). Allowing for spectral type uncertainties of $\sim$1--2 subclass, then we also use the absence of lithium (with $EW[Li] < 100$ m\AA) as a nonyouth indicator for spectral types $\le$K4 or $\ge$M6 (rejecting 19 $\le$K4 field interlopers).

\item {\em $H\alpha$ emission:} Old main sequence stars exhibit a wide range of activity levels \citep[e.g.,][]{Kafka:2006ig,West:2011zl}, and hence the presence of strong H$\alpha$ emission can not be used as a positive criterion for determining membership. However, young stars exhibit a lower bound on their H$\alpha$ emission as a function of spectral type, and this boundary has been well-determined for the similar-aged clusters IC 2602 and IC 2391 \citep[][]{Stauffer:1997ly}. We use that lower bound (which we show in Figure 6) as a nonyouth indicator (rejecting another 23 field interlopers).

\item {\em Radial velocities:} By definition, young moving groups are comoving with a very small velocity dispersion ($\sigma \sim$1 km/s for TWA; \citealt[][]{Mamajek:2005nu}), and hence the radial velocities of Tuc-Hor members should correlate very well with the projection of the group $UVW$ velocity into the line of sight. However, short-period binaries could have large velocity discrepancies due to orbital motions. SB1s in particular are impossible to distinguish from non-members in single-epoch spectroscopy. We identify candidates as members if they agree with the expected velocity of a Tuc-Hor member to within $\pm 3 \sigma$ or $\pm 3$ km/s (confirming 108 members). We label all remaining stars as likely non-members or SB1s (rejecting 20 likely field interlopers). Distinguishing the SB1s from the bona fide field stars will require additional RV measurements in the future to test for the variations denoting orbital motion.

\end{itemize}

As we summarize in Table 2, these three criteria identify 129 new Tuc-Hor members and recover 13 previously known members, while rejecting 42 confirmed field stars (based on spectroscopic youth indicators) and 20 likely field stars or SB1s (based on RVs). The new member yield for our kinematic selection process is therefore $129/191 = 67\%$. The overlap between our selection criteria provides a check on their validity. Of the 19 interlopers rejected by Li, all are also rejected by H$\alpha$ and only one has an RV consistent with membership. Conversely, of the 34 members confirmed by Li, none would be rejected by H$\alpha$ and 5 would be rejected by RVs. Of the 23 interlopers rejected by H$\alpha$, only 4 have RVs consistent with membership. The nine objects with conflicting indicators should be observed in more detail to confirm their nature, as should the 19 objects which were rejected by their RVs. Given a short-period binary frequency of $F \sim 10 \%$ for $a < 2$ AU \citep[e.g., ][]{Raghavan:2010sz}, then our 142 members should be matched by $\sim$14 SBs that could have been rejected by their discrepant RVs, much as the 34 objects confirmed by Li include 5 RV-discrepant objects that likely are SB1s.

\subsection{Comparison to Previous Surveys of Tuc-Hor}

As we discuss in Section 2.4, there are 17 well-studied members of Tuc-Hor with spectral types $\ge$K3 from \citet[][]{Torres:2006fk,Torres:2008lr}. We recovered 15 of them as candidates and re-confirmed 13 with our own spectroscopic observations. However, many other candidate members have been proposed that are not yet fully confirmed. Most programs have only identified a small number of candidates \citep[][]{Kiss:2011gf} or have concentrated on higher-mass membership \citep[][]{Zuckerman:2004fj,Torres:2006fk,Torres:2008lr,Zuckerman:2011ve}. The only large studies aimed at the low-mass population of Tuc-Hor were conducted by \citet[][]{Malo:2012fv}, who suggested 37 late-type stars (drawn from the active M dwarf sample of \citealt[][]{Riaz:2006pd}) to be candidate members, and by \citet[][]{Rodriguez:2013vn}, who suggested 58 late-type stars (identified based on GALEX excesses) to be candidate members. However, each group was only able to confirm one member based on the presence of lithium, and neither obtained RVs.

Of the 37 late-type stars identified as candidates by \citet[][]{Malo:2012fv}, we identified 23 to be candidates in our own search as well and obtained spectra for 19 of them, confirming 15 new members and rejecting 4 interlopers. Of the 14 candidates that we did not recover, two were rejected for data quality issues: one star had a spurious proper motion in UCAC3, while the other fell among a compact clustering of candidates that we attributed to a bad photographic plate, and hence rejected as a group. One candidate fell outside the RA range we considered, and two candidates had a best-fit kinematic distances of $d >$80 pc. Two candidates failed our photometric cut, falling below the main sequence for the best-fit kinematic distance. Finally, eight candidates failed our astrometric selection criterion with large values of $\Delta$ (6 with $\Delta = 10$--20 mas/yr, and 2 with $\Delta > 20$ mas/yr). \citet[][]{Schlieder:2012bv} have already shown that two of these stars are field interlopers, indicating that our more precise proper motions might be better at rejecting field stars that are only moderately discrepant from the proper motion of Tuc-Hor.\footnote{Simultaneously with our results, Malo et al. (2014) reported spectroscopic observations of 30 previously unobserved candidate Tuc-Hor members with $P_{mem} > 25\%$, mostly drawn from the Xray-selected sample of Malo et al. (2012). They confirmed 23 new members (15 of which we also confirm) and reject 7 field stars (3 of which we also reject). Our results disagree regarding the membership of 2MASS J02224418-6022476, which they confirm and we reject.}

Of the 58 late-type stars identified as candidates by \citet[][]{Rodriguez:2013vn}, we identified 35 as candidates and obtained spectra for 29 of them, confirming 26 new members and rejecting 3 interlopers. Of the 23 candidates that we did not recover as candidates, 2 were rejected for data quality issues: one had a poor reduced $\chi^2$ fit for its proper motion, and another for its SED fit.  Another 11 candidates failed our astrometric selection criterion with large values of $\Delta$ (9 with $\Delta = 10$--20 mas/yr, and 2 with $\Delta > 20$ mas/yr). Four candidates failed our photometric selection criterion, falling below the main sequence at their nominal kinematic distance by 0.1--0.5 mag. Finally, one candidate had a spectral type which was too early ($<$K3) and five candidates had spectrophotometric distances that were too large ($d \ge 80$ pc), but otherwise would have been included in our sample.

The question remains as to why \citet[][]{Malo:2012fv} and \citet[][]{Rodriguez:2013vn} did not identify our remaining 90 new Tuc-Hor members. We ultimately trace this paucity of identified candidates to their choice of input samples, which were based on ROSAT or GALEX. Most of our new members are not found in the active M dwarf census of \citet[][]{Riaz:2006pd}, and indeed, the majority do not have ROSAT counterparts at all; only 53 of our 129 previously unidentified members have an X-ray counterpart in the ROSAT All-Sky Survey within $<$30\arcsec. Most searches for nearby young stars begin with a pre-selection of candidates that are active in ROSAT \citep[][]{Shkolnik:2009lr,Shkolnik:2012ee,Malo:2012fv} or GALEX \citep[][]{Findeisen:2010mj,Shkolnik:2011qf,Rodriguez:2011ul,Rodriguez:2013vn}, but given the wide range of activity levels for even very young stars \citep[][]{Preibisch:2002qt,Preibisch:2005lc} and the limited sensitivity of these surveys for stars at $>$25 pc, then this pre-selection must be pursued with great caution or ultimately might need to be abandoned in favor of purely kinematic criteria (as in our program). We address the role of activity selection with GALEX data in more detail in Section 6.7.

\section{The Population Statistics of Tucana-Horologium}

 \begin{figure*}
 \epsscale{1.0}
 \plotone{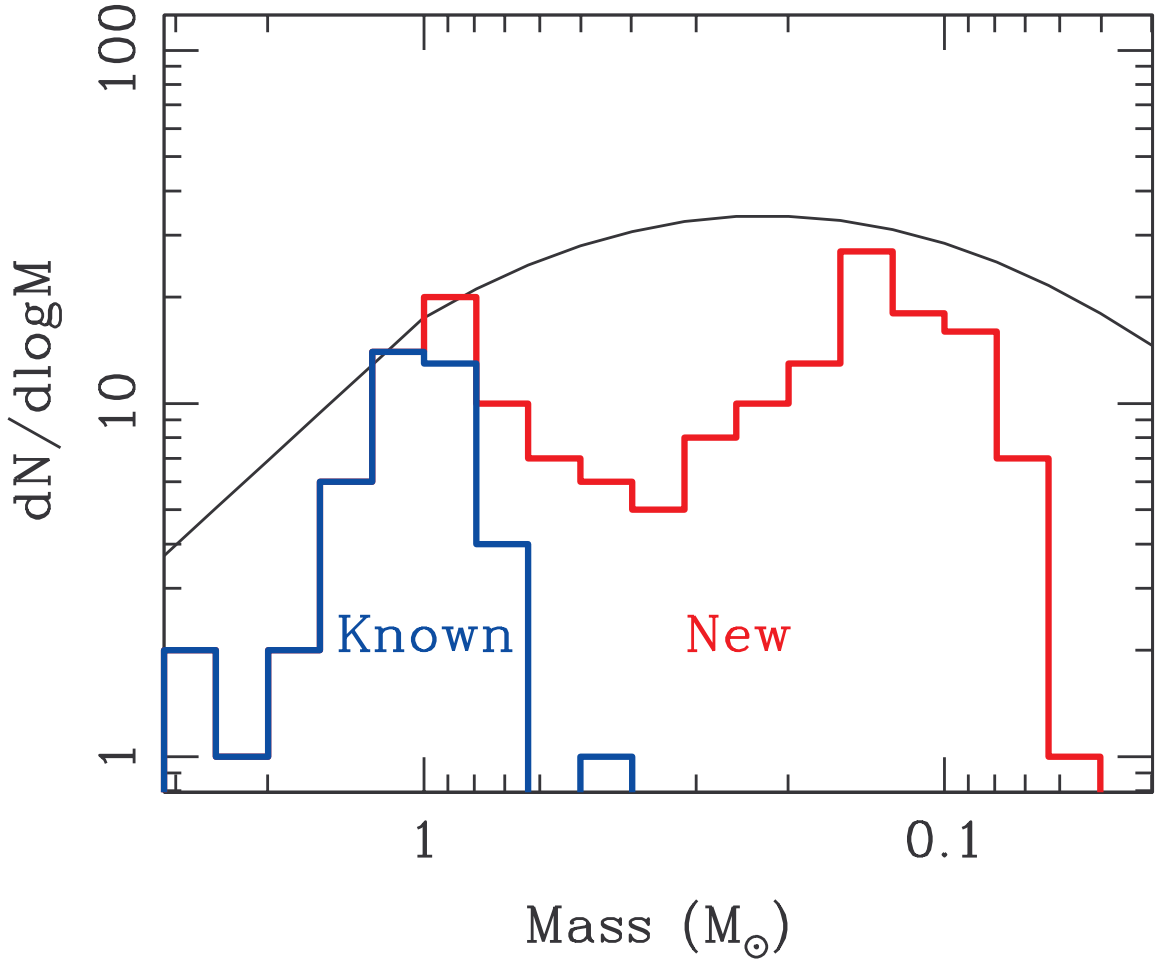}
 \caption{The mass function for Tuc-Hor, showing previously known members (blue) and our newly-identified members (red). We also show the Salpeter+Chabrier IMF (black line), normalized for the two bins around $M = 1 M_{\odot}$. Even if we assume that the solar-mass membership is complete and accurate, then there are a significant number of M dwarfs remaining to be discovered, as well as virtually all of the brown dwarfs.}
  \end{figure*}

 \begin{figure*}
 \epsscale{1.0}
 \plotone{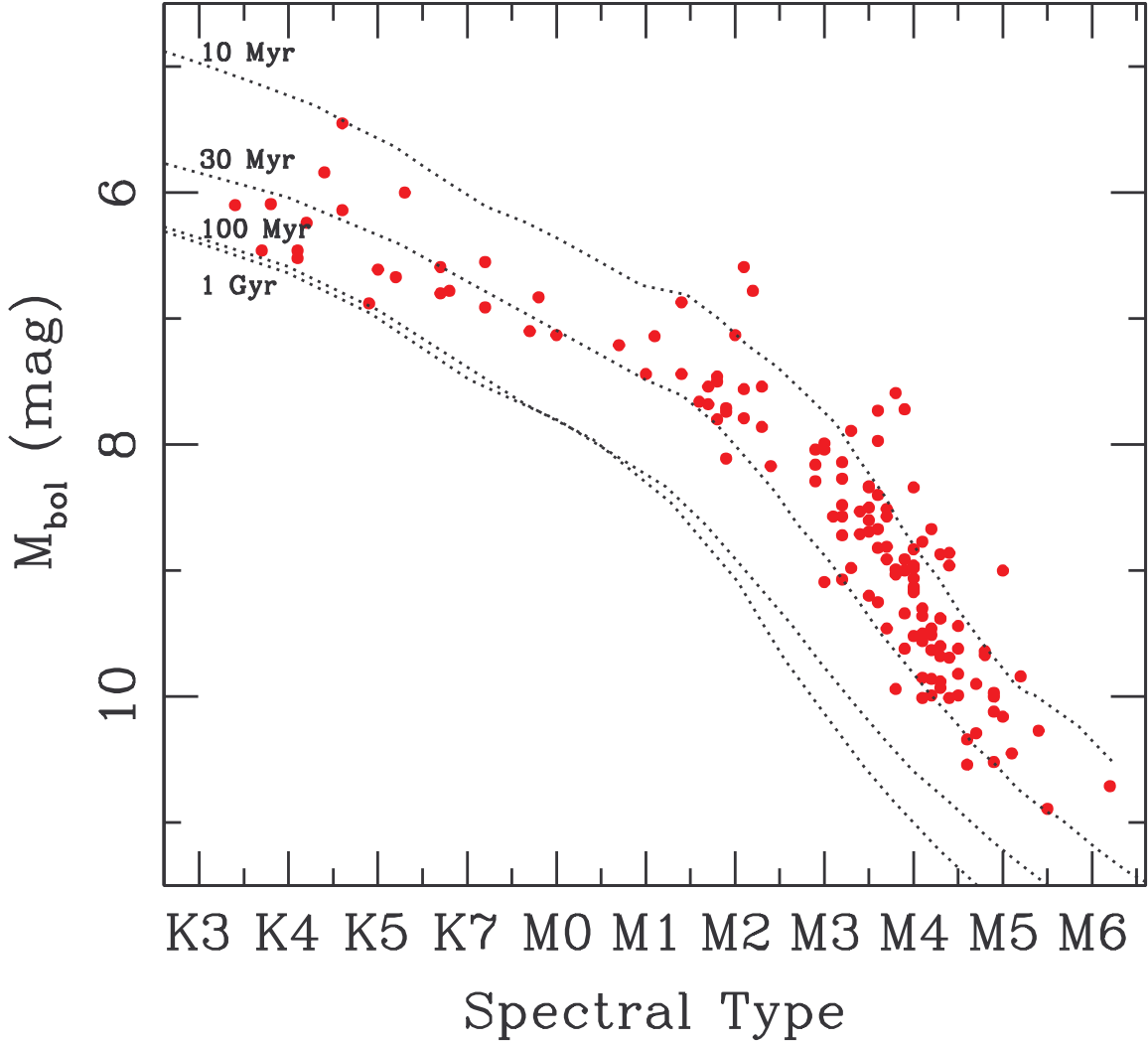}
 \caption{HR diagram for the 142 Tuc-Hor members that we observed. We also show the isochrones of \citet[][]{Baraffe:1998yo} for ages of 10 Myr, 30 Myr, 100 Myr, and 1 Gyr; the model temperatures are converted to spectral types using the dwarf temperature sequence of \citet[][]{Golimowski:2004fr}. The members of Tuc-Hor fall in between the 10 Myr and 30 Myr isochrones, implying an isochronal age of $\tau \sim 20$ Myr. Using the young-star temperature sequence of \citet[][]{Luhman:2003pb} yields an older age of $\tau \sim 30$ Myr; the dwarf and young-star temperature scales of \citet[][]{Pecaut:2013zr} are both intermediate between these extremes.}
  \end{figure*}

 \begin{figure*}
 \epsscale{0.36}
 \plotone{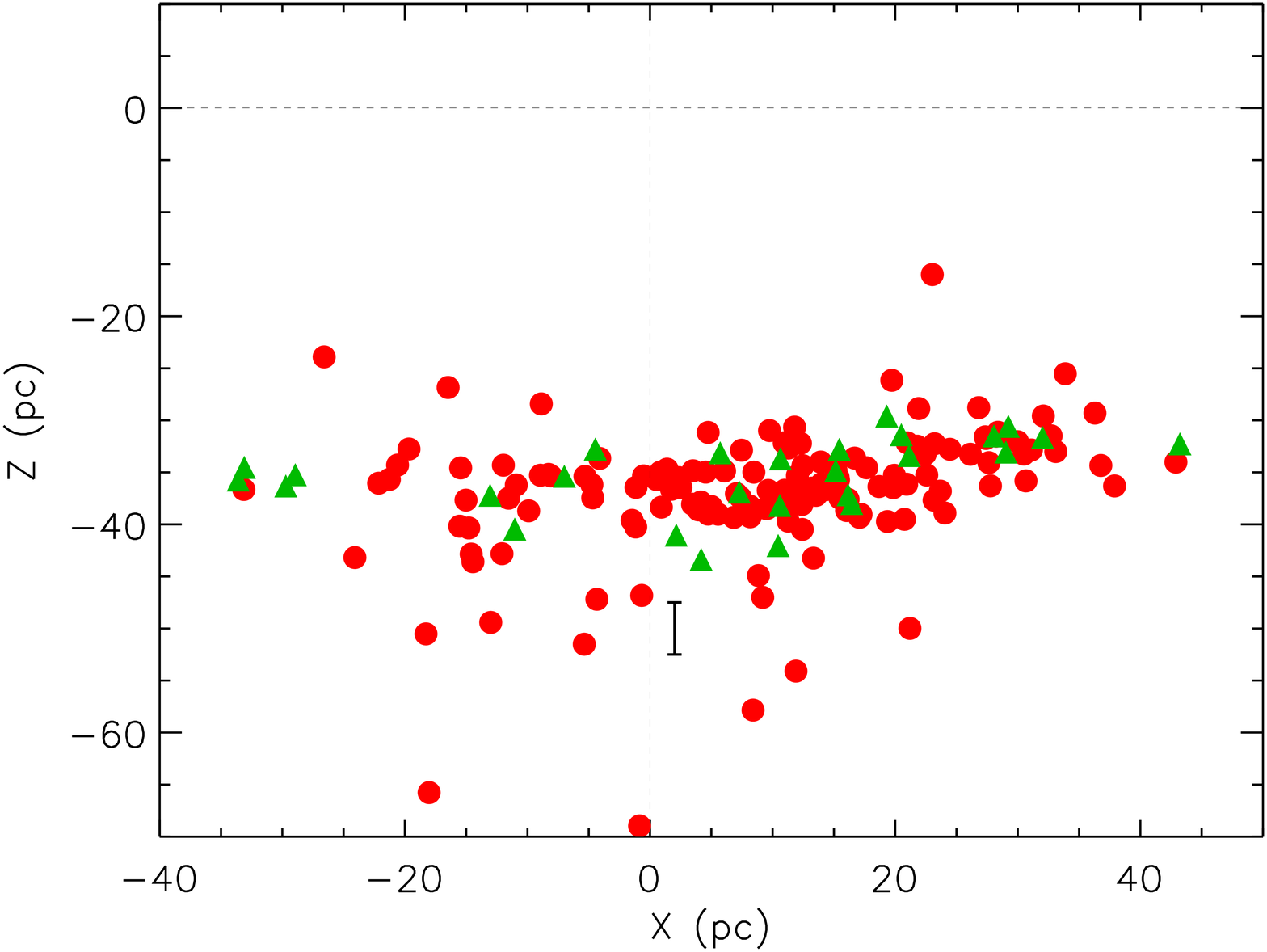}
 \epsscale{0.36}
 \plotone{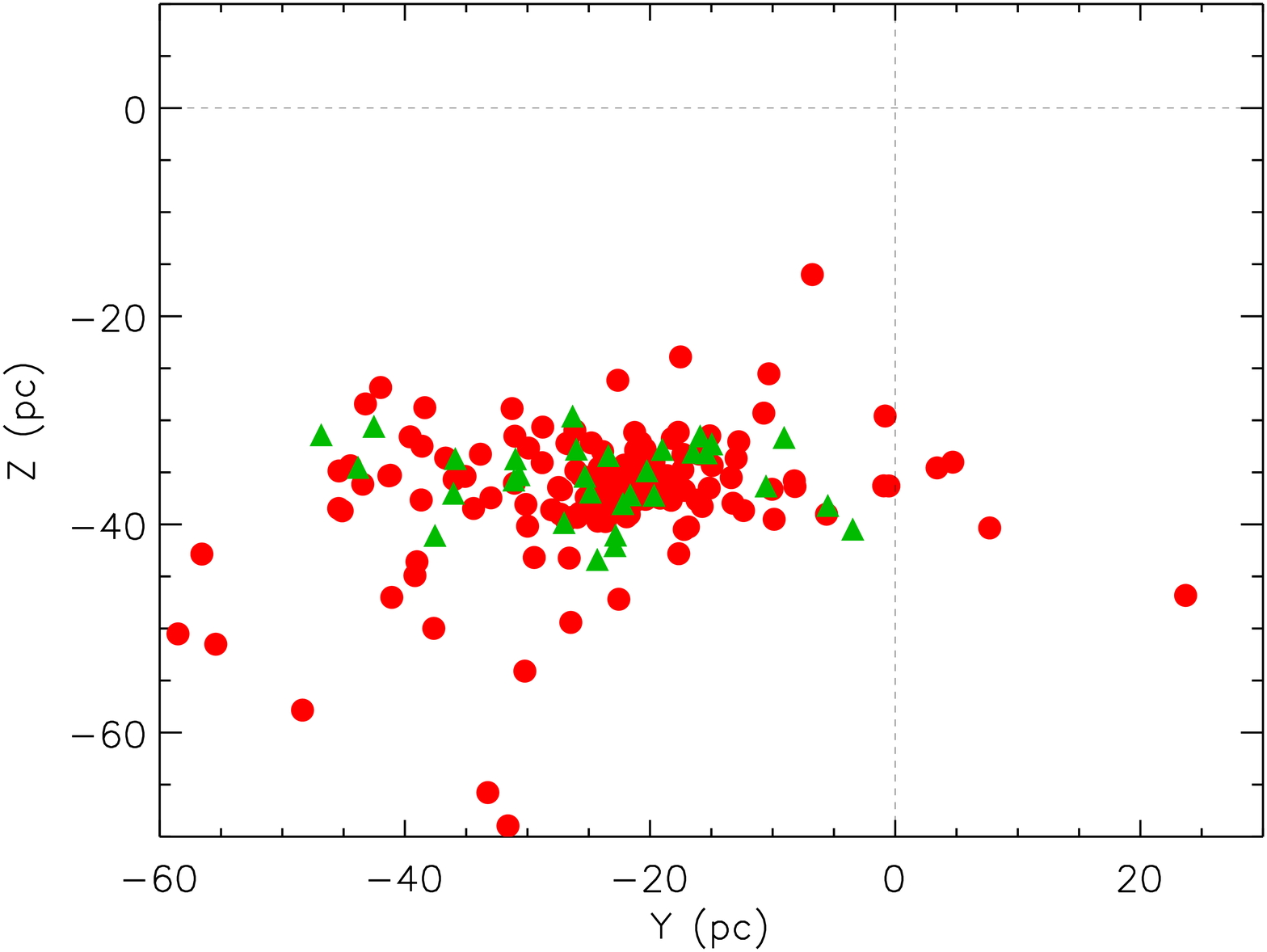}
 \epsscale{0.36}
 \plotone{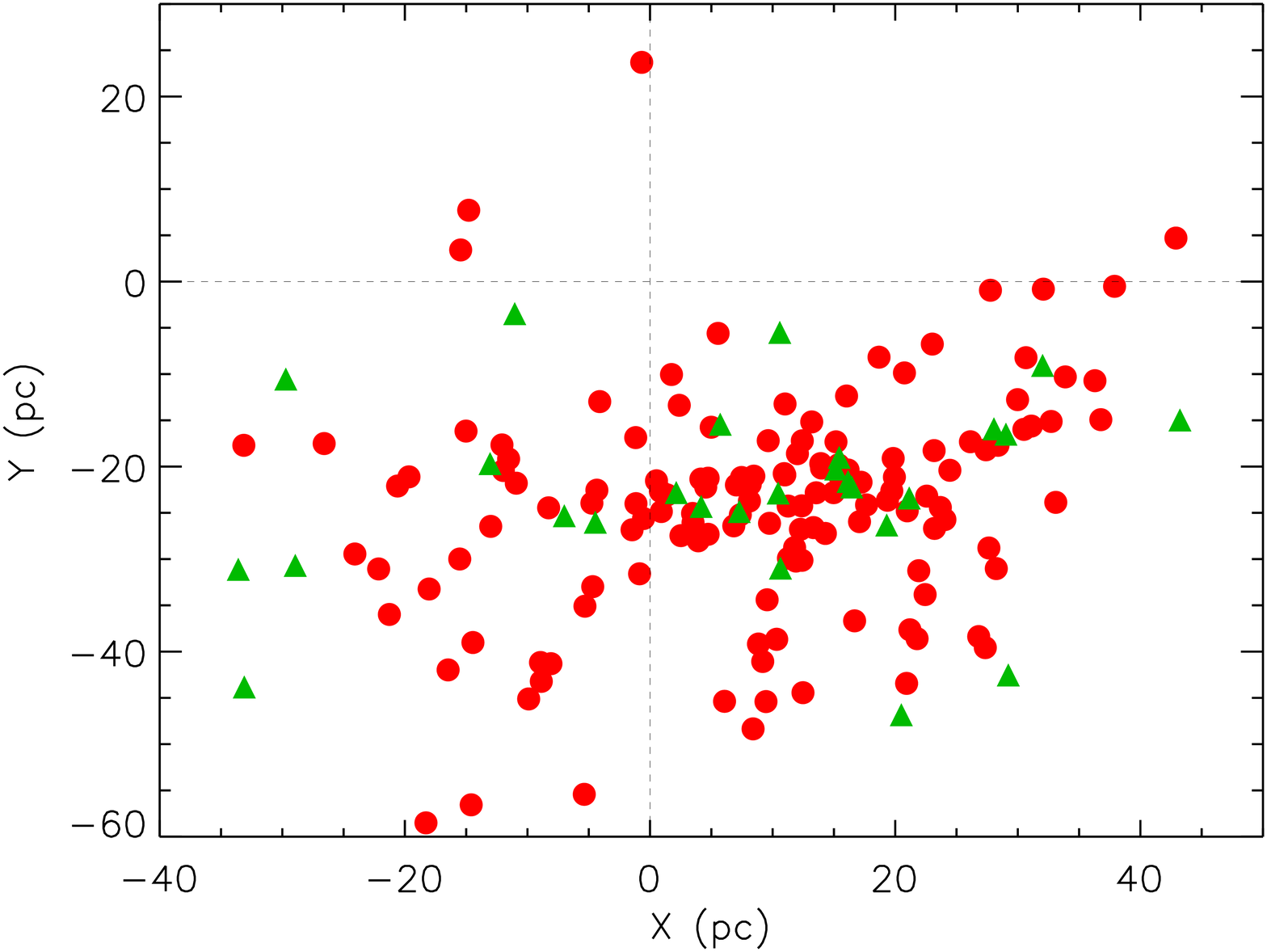}
 \caption{The XYZ spatial distribution of the 142 K3--M6 Tuc-Hor members that we observed (red circles)) and the 31 known members with SpT$<$K3 (green triangles; \citealt[][]{Torres:2008lr}).  The (X,Z) and (Y,Z) plots clearly show that the distribution is narrow in $Z$, while the (X,Y) plot shows that it is extended in both of those axes. Given the location of Tuc-Hor on the celestial sphere, $Z$ primarily indicates distance, while $X$ and $Y$ primarily indicate sky position. As can be seen in Figure 2, there are confirmed members across the entire range of sky that we observed, and hence we can not comment on the extent of this planar structure in $X$ or $Y$. We also show a characteristic 1$\sigma$ error bar in the (X,Z) plot to demonstrate the typical uncertainty in kinematic distances. Given the scatter of $\sim$5 mas/yr in measurements of $\Delta$, then the proper motions should be uncertain by that amount, yielding kinematic distance uncertainties of $\pm$5\% or $\pm$2--3 pc. (Away from $X=0$ pc, the error bars will be rotating to point toward the origin.)}
  \end{figure*}
 
 \begin{figure*}
 \epsscale{1.0}
 \plotone{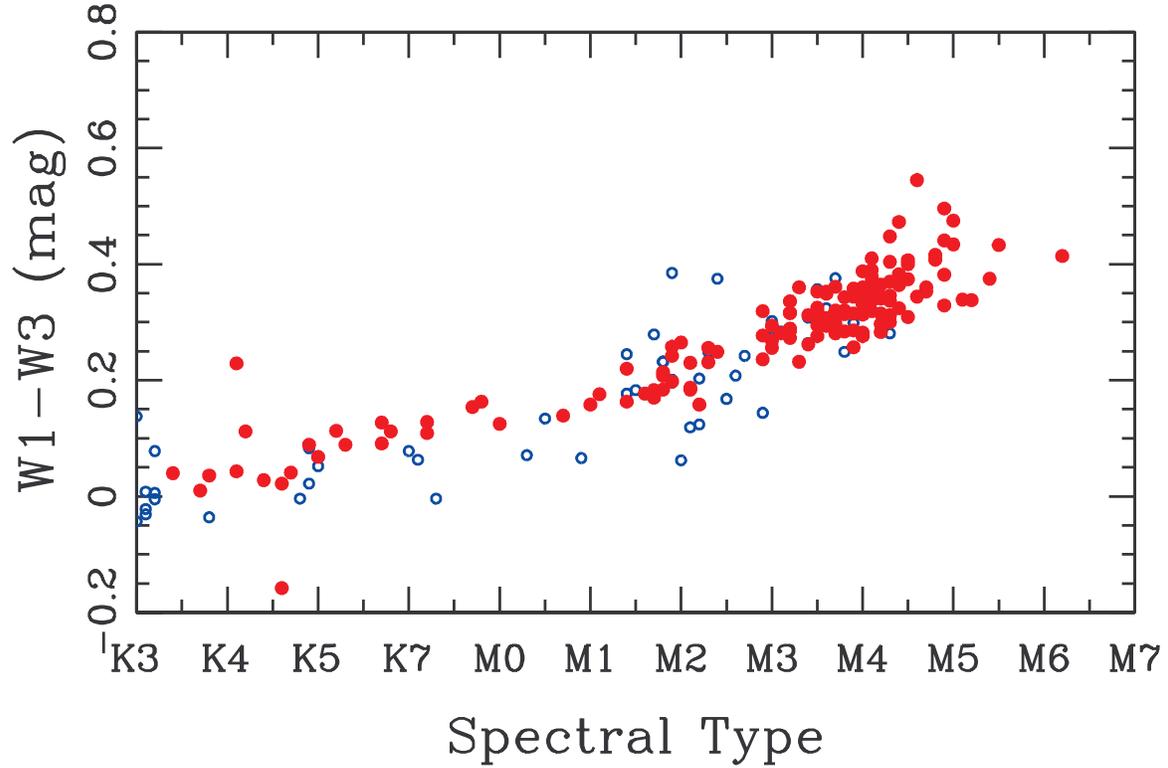}
 \caption{WISE $W1-W3$ color as a function of (spectroscopically determined) spectral type for our observed Tuc-Hor members (red filled circles) and apparent field interlopers (blue open circles). According to the criteria suggested by Luhman \& Mamajek (2012), all of our targets are Class III (diskless) sources.} 
  \end{figure*}

 \begin{figure*}
 \epsscale{1.0}
 \plotone{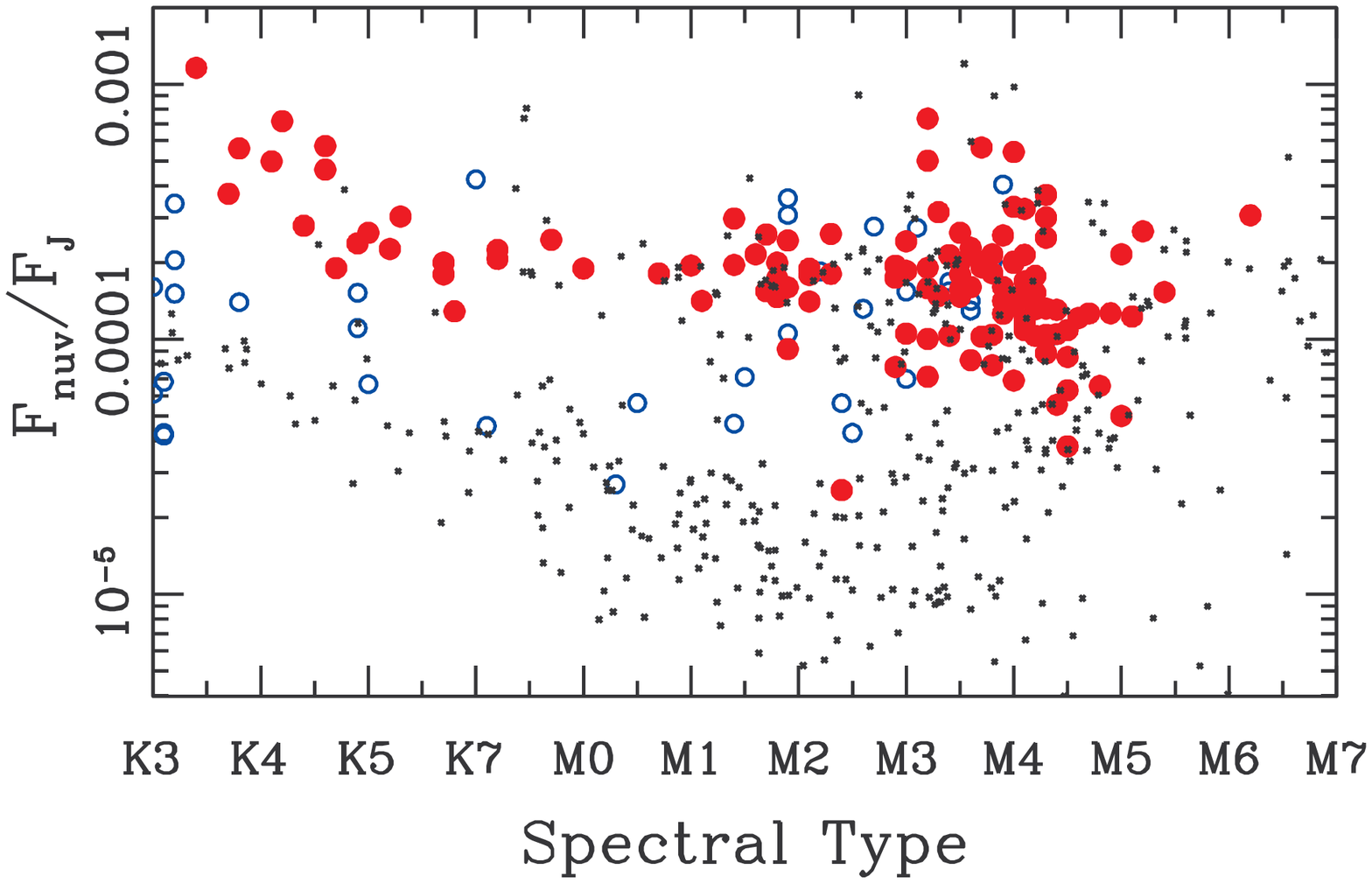}
 \caption{Fractional near-UV flux density as a function of (spectroscopically determined) spectral type for our observed Tuc-Hor members (filled red circles), apparent field interlopers (blue open circles), and the K-M stars from the NStars 25 pc sample (small black crosses; \citealt[][]{Reid:2007dp}). As we discuss in Section 6.7, the young star sequence can be clearly divided from most (presumably older) field stars for K2-M2 stars, but the sequences increasingly overlap for spectral types later than M2. The star that is comoving in RV and sits well below the Tuc-Hor sequence in fractional UV flux density (2MASS J04133314-5231586) could be a field interloper. Its H$\alpha$ emission strength is consistent with the member sequence, but on the lower edge.}
  \end{figure*}

\subsection{Mass Function}

As we show in Figure 2 and discuss further in Section 6.4, it is unlikely that our survey is spatially complete, and it is not clear whether our survey even encompasses the same spatial volume as the surveys that identified the known higher-mass members. As a result, any mass function for the region must be considered extremely preliminary. However, plotting the mass function of the members discovered to date still can be very illustrative. Given the strong evidence that the IMF is universal for most young populations in the solar neighborhood (e.g., \citealt[][]{Bastian:2010wb}  and references therein), then a comparison of Tuc-Hor to the standard IMF can demonstrate which mass ranges of members are still incomplete.

In Figure 8, we show the mass function ($dN/d\log M$) for the known members of Tuc-Hor and our newly-discovered members. We inferred masses from the observed spectral types using the mass-$T_{eff}$ relations of \citet[][]{Baraffe:1998yo} and \citet[][]{Siess:2000ce} (for $\le$1.4 $M_{\odot}$ and $>$1.4 $M_{\odot}$ members, respectively), combined with the dwarf $T_{eff}-SpT$ temperature scale of \citet[][]{Schmidt-Kaler:1992ab}. We also show the Salpeter IMF \citep[][]{Salpeter:1955bh} for $>$1 $M_{\odot}$ stars and the Chabrier IMF \citep[][]{Chabrier:2003lq} for $\le$1 $M_{\odot}$ stars, normalized to the observed number of Tuc-Hor members in the two bins straddling 1 $M_{\odot}$.

The paucity of stars at $M = 0.2$--0.7 $M_{\odot}$ (SpT$=$ M0--M3) indicates that our survey is indeed incomplete. Given the apparently higher rate of completeness for VLM stars in the $M = 0.07$--0.2 $M_{\odot}$ range, then mere spatial incompleteness appears unlikely, given that we observed a similar fraction of candidates in both mass ranges. We instead speculate that this paucity of early-M stars might result from errors in our SED templates. As we discussed in Section 2.3, it appears that M0--M1 stars are being pulled to a spectral type of $\sim$K7.5. The inferred $m_{bol}$ (which is effectively set by the sum of the observed flux in all filters) would not change, and hence any candidates in the M0--M1 range would appear to be underluminous K7.5 stars and would tend to fall under our photometric selection criterion. A future reanalysis of the entire sky with updated SED templates should demonstrate if this is the case, yielding the missing candidates.

\subsection{HR Diagram and Isochronal Age}

In Figure 9, we show an HR diagram for the confirmed members, plotting $M_{bol}$ as a function of spectroscopic spectral type. The absolute $M_{bol}$ for each star is calculated from the apparent $m_{bol}$ derived from the SED fit and the kinematic distance modulus derived from the proper motion. We also show the 10 Myr, 30 Myr, 100 Myr, and 1 Gyr models of \citet[][]{Baraffe:1998yo}, as derived with a convective scale length of 1.9 times the pressure scale height. We converted the model $T_{eff}$ values to spectral types using the dwarf temperature sequence of \citet[][]{Schmidt-Kaler:1992ab} for $\le$M0 stars and the dwarf sequence of \citet[][]{Golimowski:2004fr} for $\ge$M1 stars. If we use the young-star temperature sequence of \citet[][]{Luhman:2003pb} for $\ge$M1, the sequence is shifted $\sim$0.5 subclass later for a given mass or temperature. The dwarf and young-star temperature scales of \citet[][]{Pecaut:2013zr} both fall between these limits.

Using the dwarf sequence, we find that the median isochronal age for Tuc-Hor is $\tau \sim 20$ Myr. For the young-star temperature sequence, the age is shifted to $\tau \sim 30$ Myr. The results of \citet[][]{Pecaut:2013zr} suggest that the young-star temperature sequence might be more appropriate even for intermediate-age populations like Tuc-Hor, and hence it remains unclear which sequence should be preferred. As we discuss below, both of these ages are younger than the lithium depletion boundary age of $\tau \sim 40$ Myr. This trend is consistent with the results of \citet[][]{Pecaut:2012dp}, who use several other age diagnostics to determine that the Upper Scorpius subgroup of the Sco-Cen OB association might be a factor of $\sim$2 older ($\tau \sim 11$ Myr) than its traditionally accepted isochronal age for low-mass members ($\tau \sim 5$ Myr; \citealt[][]{Preibisch:2002qt}). \citet[][]{Binks:2013aa} also have demonstrated a similar discrepancy for the BPMG.

\subsection{Lithium Depletion Age}

As we discussed in Sections 4.3 and 5.1, lithium depletion is a key indicator of age for low-mass stars, being depleted on timescales of $\la$10 Myr for early-M stars and $\sim$100 Myr for stars across the full range of spectral types we consider. Lithium also can be used to age-date stellar populations as a whole, placing them in a relative age sequence based on the location of the lithium depletion boundaries (for both late-K stars and early-M stars) as a function of spectral type or absolute magnitude. The depletion of lithium in K stars has long been used to age-date populations \citep[][]{King:2000cr}, but age-dating with mid-M stars is less widespread because these low-mass members have been more difficult to identify.

The precise location of the boundary can be difficult to quantify, as the observed properties of moving group members can be blurred by observational uncertainties (most notably in distance or spectral type) or astrophysical effects (unresolved binarity, rotation, or genuine age spreads). We therefore have quantified the location of the lithium depletion boundaries by identifying the limit where equal numbers of lithium-depleted and lithium-bearing stars encroach onto the opposite side of the boundary. To avoid a bias from our use of lithium as a membership indicator, we only consider those lithium-bearing stars that were also selected based on RVs. 

Using this definition, we find that the late-K lithium boundary is at (spectroscopically determined) SpTs of K5.5 $\pm$0.3 (where two earlier members are lithium depleted and another two later members are lithium-bearing). The mid-M boundary is at M4.5 $\pm$ 0.3 (with 6 members violating each side of the boundary). The corresponding boundaries for SED-fot SpTs are K7.6 $\pm$ 0.6 (4 members) and M4.7 $\pm$ 0.7 (8 members). For absolute bolometric luminosities ($M_{bol} = m_{bol} + DM_{kin}$), the boundaries are at $M_{bol} = 6.64 \pm 0.20$ (3 members) and $M_{bol} = 9.89 \pm 0.10$ (5 members). Finally, for absolute $K_s$ magnitudes, the boundaries are at $M_{Ks} = 4.33 \pm 0.15$ (3 members) and $M_{Ks} = 7.12 \pm 0.16$ (5 members). In each case, we estimate the uncertainty from the range encompassing a number of non-encroaching objects equal to the number of encroaching objects. For absolute bolometric luminosities ($M_{bol} = m_{bol} + DM_{kin}$), the boundaries are at $M_{bol} = 6.64 \pm 0.20$ (3 members) and $M_{bol} = 9.89 \pm 0.10$ (5 members). Finally, for absolute $K_s$ magnitudes, the boundaries are at $M_{Ks} = 4.33 \pm 0.15$ (3 members) and $M_{Ks} = 7.12 \pm 0.16$ (5 members). In each case, we estimate the uncertainty from the range encompassing a number of non-encroaching objects equal to the number of encroaching objects.

The late-K depletion boundary only changes subtly at ages of $\ga$10 Myr, and hence it is challenging to construct an unambiguous sequence. \citet[][]{Randich:2001lr} studied the $\sim$50 Myr clusters IC 2602 and IC 2391, and for the same boundary definition as described above, they found it to fall at $T_{eff} = 4025$ K (or SpT = K7.1 from the temperature scale of \citealt[][]{Schmidt-Kaler:1992ab}). \citet[][]{Balachandran:2011rr} found that in the older ($\sim$75 Myr) $\alpha$ Per cluster, the boundary falls at $T_{eff} = 4735$ K (SpT = K3.2) with 3 interlopers. In the canonically $\sim$125 Myr Pleiades cluster, \citet[][]{King:2000cr} found the boundary to lie at $T_{eff} = 4420$ K (SpT = K4.7). For these and many other clusters, a more diagnostic estimate can be derived from examining the full sequence of $EW[Li]$ versus SpT for FGK stars, as lithium depletion occurs gradually across this full range. However, our census only adds a modest number of stars with SpT earlier than K5, so we refer the reader to a comprehensive analysis of the known higher-mass members by \citet[][]{Mentuch:2008vn} and \citet[][]{da-Silva:2009wd}.

The evolution of the mid-M lithium depletion boundary is more unambiguous due to the large dynamic range of $M_{bol}$ over which it varies in relevant age scales. \citet[][]{Barrado-y-Navascues:2004nx} reported a boundary at $M_{bol} = 10.24 \pm 0.15$ in IC 2391, as well as updating the results of \citet[][]{Stauffer:1998mz,Stauffer:1999eu} for $\alpha$ Per ($M_{bol} = 11.31 \pm 0.15$) and the Pleiades ($M_{bol} = 12.14 \pm 0.15$). \citet[][]{Dobbie:2010oq} reported the boundary to fall at $M_K = 7.37 \pm 0.20$ for IC 2602, which is equivalent to $M_{bol} = 10.22 \pm 0.22$ given that $BC_K = 2.85 \pm 0.10$ for M5.0-M5.5 stars \citep[][]{Kraus:2007mz}. \citet[][]{Cargile:2010rr} reported a boundary for Blanco 1 at $M_{bol} = 11.99 \pm 0.30$. Finally, a very recent measurement for new members of the BPMG by \citet[][]{Binks:2013aa} found the lithium depletion bound to fall at $M_{bol} = 8.3 \pm 0.5$, corresponding to an age of $\tau = 21 \pm 4$ Myr.

As we summarize in Table 5, our measurement of the mid-M lithium depletion boundary for Tuc-Hor ($M_{bol} = 9.89 \pm 0.10$ or $M_{Ks} = 7.12 \pm 0.16$) indicates an age consistent with that of IC 2391 and IC 2602 ($\sim$45 Myr), clearly older than BPMG ($\tau = 21$ Myr) and clearly younger than the other reference populations. The evolutionary models of \citet[][]{Baraffe:1998yo} and \citet[][]{DAntona:1997li} imply lithium depletion ages of $41 \pm 2$ Myr and $38 \pm 2$ Myr respectively, where the uncertainties reflect only the uncertainty in the boundary location. The real error budget is most likely dominated by the uncalibrated nature of the models themselves.

\subsection{The Spatial Structure of Tuc-Hor}

In Figure 10, we show the $XYZ$ spatial distributions for our observed members and for the SpT$<$K3 members that were previously known \citep[][]{Torres:2008lr}. The two distributions broadly match and demonstrate that the main body of Tuc-Hor is compact in $Z$, with a median value of $Z = -36$ pc and a total extent of $\pm$5 pc for all but a few extreme outlying members. In contrast, the distribution is very broad in the ($X$,$Y$) plane. A similar spatial distribution can be seen in the activity-selected candidates reported by \citet[][]{Rodriguez:2013vn}. Inspection of Figure 2 shows that we did not observe the candidates which would fall near the edges of the ($X$,$Y$) panel of Figure 10, and hence we can not comment on the total extent in this plane. Finally, a visually recognizable overdensity is located at ($XYZ$) $\sim$ (+10,-25,-35) pc, corresponding to the traditionally identified ``core'' of Tuc-Hor which has on-sky coordinates of ($\alpha$,$\delta$) $\sim$ (2$^h$,-60$^o$) and which is equally recognizable in Figure 2.

As for other moving groups, like TWA (e.g., \citealt[][]{Weinberger:2012qy}), our results demonstrate that the Tuc-Hor population is not distributed in an ellipsoid. TWA shows broadly filametary structure, while Tuc-Hor more closely resembles a sheet. These populations are too young to have been distorted by the Milky Way's tidal field, having only existed for $\la$1/8 of a galactic orbit, and hence this geometry must trace a combination of the primordial molecular cloud structure and specific forces (such as interactions with molecular clouds) that induced non-spherical velocity dispersions. If the former effect dominates, then it would indicate that geometric analyses are of limited use for determining trace-back ages, and more generally that these moving groups formed in a distributed manner \citep[similar to Taurus-Auriga; ][]{Simon:1997er,Kraus:2008fr} rather than as compact clusters that have since become unbound \citep[more akin to $\eta$ Cha; ][]{Murphy:2010lr}.

Finally, the small extent in $Z$ places a strong constraint on the internal velocity dispersion. Assuming that Tuc-Hor formed in a sheet with zero thickness in the $Z$ axis, then a typical member has moved by $<$5 pc in the moving group's lifetime of $\sim$30 Myr. The corresponding one-dimensional velocity dispersion for individual members (or for substructures with typical scales of $\la$5 pc) is only $\sigma_v \sim 160$ m/s. This velocity dispersion is comparable to the small-scale velocity dispersion seen in Taurus-Auriga \citep[][]{Kraus:2008fr}, further supporting Tuc-Hor's origin as a dynamically quiet T association.

\subsection{The UVW Velocity and Dispersion of Tuc-Hor}

For any individual member of a stellar population, the radial velocity $v_{rad}$ indicates the one-dimensional projection of the population's space velocity $v_{UVW}$ onto the line of sight $d$ toward that member ($\vec{v}_{rad} = \vec{v}_{UVW} \cdot \hat{d}$). When a stellar population subtends a large solid angle of the sky, then the radial velocities of its members collectively trace a wide range of such projections, and hence can be used to tomographically reconstruct the full three-dimensional value of $v_{UVW}$. This geometric reconstruction can place extremely tight constraints on the space motion, since the measurement of $v_{rad}$ is limited only by the intrinsic velocity dispersion of the cluster and the instrumental precision, whereas full $v_{UVW}$ measurements for individual stars are generally limited by the precision of the proper motion and by the distance (which is needed to convert the proper motion from an angular velocity to a spatial velocity).

For Tuc-Hor, we computed this tomographic reconstruction by conducting a grid search of $UVW$ velocities, finding the mean $UVW$ velocity that minimizes the $\chi^2$ of the fit and determining confidence intervals in the $\chi^2$ surface around that minimum. For this calculation, we used 65 stars that have observational uncertainties $\sigma_{vrad} < 1$ km/s (to reject fast rotators and other stars with noisy measurements) and for which their velocities agree with the expected value to within $<$3 km/s (to reject spectroscopic binaries). The resulting space motion at the minimum in the $\chi^2$ surface is $v_{UVW} = $(-10.6, -21.0, -2.1) km/s, with 1$\sigma$ uncertainties on each dimension of $\pm$0.2 km/s. The reduced $\chi^2$ value for our best-fit value of $v_{UVW}$ is $\chi^2_{\nu} = 7.4$ (with 62 degrees of freedom), indicating that the velocity dispersion is significantly resolved compared to our estimated uncertainties on the RVs. We therefore increased the uncertainties by a factor of $\sqrt{7.4}$ before calculating the 1$\sigma$ uncertainty on the mean $v_{UVW}$. Our value for the mean $v_{UVW}$ is very close to the canonical velocity of $v_{UVW} = $(-9.9, -20.9, -1.4) km/s \citep[][]{Torres:2008lr}, but is considerably more precise, even after increasing our RV uncertainties so that $\chi_{\nu}=1$.

If we compare the expected radial velocity for each star ($\vec{v}_{rad} = \vec{v}_{UVW} \cdot \hat{d}$) to the measured values, we find that the scatter of our measured RVs about the best-fit values is $\pm$1.1 km/s. This scatter could result from either the noise floor of our RV measurements (which we have ruled out via multiple observations of a subset of standard stars; Section 3) or the intrinsic velocity dispersion across all of Tuc-Hor. The physical arguments from the previous section motivate a low velocity dispersion on small angular scales ($\sigma_v \sim 160$ m/s on scales of $\la 5$ pc). However, the total extent of Tuc-Hor is large ($\ga$50 pc), and the velocity dispersion in molecular clouds (and the resulting stellar populations) should increase on large angular scales by $v \propto d^{0.5}$ \citep[][]{Larson:1981qe}. We therefore expect the velocity dispersion on a spatial scale of 50 pc to be $\sim$3 times larger than the velocity dispersion on a spatial scale of 5 pc, which would account for part of the larger dispersion in our RV measurements.

\subsection{Disk Frequency}

Several members of the BPMG and TWA moving groups are known to host disks, either optically thick protoplanetary disks (as for TW Hya) or optically thin debris disks (as for $\beta$ Pic). Since the Tuc-Hor moving group is only moderately older, it is plausible that some Tuc-Hor members might also host disks. To search for these disks, we cross-referenced our list of 143 K3-M6 members of Tuc-Hor with the all-sky catalog of the Wide-Field Infrared Survey Explorer (WISE), which observed the full sky in 4 bands spanning 3.4--22 $\mu$m. In all cases, our targets are sufficiently bright to be detected in the $W1$ (3.4 $\mu$m), $W2$ (4.5 $\mu$m), and $W3$ (12 $\mu$m) bands; only 24 were detected in the $W4$ (22 $\mu$m) band.

In Figure 11, we show a plot of $W1-W3$ as a function of spectral type. Based on the criteria suggested by \citet[][]{Luhman:2012wd}, all targets are Class III (diskless) sources with $W1-W3 < 1$. The 24 stars that were detected in W4 also are consistent with photospheric colors ($W1 - W4 < 0.5$). We therefore conclude that the number of stars hosting significant quantities of warm circumstellar dust in this spectral type range is $F < 0.7\%$, with $F < 0.8\%$ for M0.0--M6.0 stars and $F < 5\%$ for K3.0--K7.9 stars.

\subsection{Selection of Young Stars with GALEX}

Stellar activity is known to be an indicator of youth \citep[][]{Preibisch:2002qt,Preibisch:2005lc}, so X-ray and UV all-sky surveys are well-suited to finding active young stars. However,  the  \rosat\ X-ray catalogs (e.g.~\citealt[][]{Voges:1999kl}) are generally limited to the nearest, earliest-type M dwarfs, since their luminosities  are $\sim$10--300$\times$ lower than solar-type stars. As we discussed in Section 5.2, this insensitivity to low-activity M dwarfs places a fundamental limit on \rosat\ selection of young stars.

Several teams have shown the higher completeness of using UV wavelengths to search for young M dwarfs \citep[][]{Findeisen:2010mj,Shkolnik:2011qf,Rodriguez:2011ul,Rodriguez:2013vn}, making the NASA Galaxy Evolution Explorer (\galex; \citealt[][]{Martin:2005ai}) a useful resource with which to expand the young low-mass census. The \galex\ satellite has imaged most of the sky simultaneously in two bands: near-UV (NUV; 1750--2750 \AA) and far-UV (FUV; 1350--1750 \AA), with angular resolutions of 5\arcsec\, and 6.5\arcsec. The full description of the instrumental performance is presented by \citet[][]{Morrissey:2005tg}. The \galex\ mission produced a relatively shallow All-sky Imaging Survey as well as several deeper surveys which collectively cover $\approx$3/4 of the sky. The NUV and FUV fluxes and magnitudes were produced by the standard \galex\ Data Analysis Pipeline (ver.~4.0) operated at the Caltech Science Operations Center \citep[][]{Morrissey:2005tg}\footnote{The data presented in this paper made use of the seventh data release (GR7). See details at http://www.galex.caltech.edu/researcher/techdoc-ch2.html.}  and archived at the Mikulski Archive at the Space Telescope Science Institute (MAST).

Previous searches for low-mass YMG members using \galex\ used color and proper motion cuts coupled with NUV and/or FUV selection criteria (\citealt[][]{Shkolnik:2011qf} and \citealt[][]{Rodriguez:2011ul} for TWA and \citealt[][]{Rodriguez:2013vn} for Tuc-Hor) before acquiring optical spectra for candidate confirmation. In this work, we did not use \galex\ to pre-select UV-active candidates, in order to avoid any bias against low-activity members. Our purely kinematic and color-magnitude selection procedure now allows us to test the efficiency and completeness of \galex-selected surveys for young stars.

In \citet[][]{Shkolnik:2011qf} we used the NStars 25-pc census ($\approx$1500 M dwarfs; \citealt[][]{Reid:2007dp}) to calibrate our \galex\ selection criteria. Namely, we identified young M dwarfs  ($<300$ Myr) as having fractional flux densities $F_{NUV}/F_J$ $>$ 10$^{-4}$ and, if detected, $F_{FUV}/F_J$ $>$ 10$^{-5}$, while the quiescent emission of old stars (those with $F_{FUV}/F_J$ $<$ 10$^{-5}$ and no \rosat\ detection) traces out a clear sequence which lies below the young, \rosat\ detected M dwarfs.  For stars earlier than K2, the Tuc-Hor and field sequences converge, and hence the \galex\ NUV cut is not a distinguishing criterion. However, in these cases, the FUV cut of $F_{FUV}/F_J$ $>$ 10$^{-5}$ can instead distinguish young stars.  

In Figure 12, we plot the GALEX NUV flux density (normalized to $J$ band flux density, $F_{NUV}/F_J$) as a function of spectral type for our Tuc-Hor candidates. Of the 204 candidates, 166 (80\%) had a \galex\ counterpart in the NUV bandpass within $<$10\arcsec, while 26 were not observed by GALEX; 138 of the 204 candidates (69\%) were observed and lie above the $F_{NUV}/F_J > 10^{-4}$ threshold used by \citet[][]{Shkolnik:2011qf}. We found that 107 of the high-NUV emitters (78\%) are confirmed members, while the remaining 31 (22\%) presumably are either Tuc-Hor SBs that had discrepant RVs in our observation epoch, other young stars ($\tau \la 300$ Myr), or old field SBs which are tidally-locked into fast rotation (and hence high activity).

Only 80 of our candidates were detected in the FUV bandpass, with all but one having $F_{FUV}/F_J$ $>$ 10$^{-5}$. We found that 68/80 are confirmed Tuc-Hor members. We therefore find that the FUV criterion works well for the more massive stars in Tuc-Hor, but fails at $d \ga 40$ pc for a significant fraction of young stars with SpT$>$M2, where they are too faint to be detected in the FUV.

Of the 142 Tuc-Hor members observed in this paper, 13 were not observed by \galex\, 5 were observed and not detected, 3 were not identified due to confusion with brighter neighbors, and 14 would have been rejected using the $F_{NUV}/F_J$ $>$ 10$^{-4}$ criterion. Therefore, had we pre-selected candidates using the NUV \galex\ criterion from \citet[][]{Shkolnik:2011qf}, we would have identified 107 of 142 (77\%) of the confirmed members as Tuc-Hor candidates. Had we first applied the NUV criterion prior to collecting additional data, we would have needed spectra of 138 stars to confirm 107 new members, yielding a confirmation efficiency of 78\%.  Without the NUV criterion, we needed 204 spectra to confirm 142 members, yielding a confirmation efficiency of 70\%. Therefore, adding the \galex\ criterion to the candidate-selection process discussed in Section~2 is somewhat more efficient, but it limits the search to $\approx$75\% of the existing members due to incomplete sky coverage and the intrinsic spread in the intrinsic $NUV$ excesses of young stars.

Finally, we can use the results of our kinematic+CMD selection procedure (which is unbiased toward stellar activity) to set a new SpT-dependent lower envelope for NUV fluxes of $\tau = 40$ Myr young stars. We find that for K3--M2 stars, the lower envelope is defined by a linear relation connecting (SpT=K3, $F_{NUV}/F_J=2 \times 10^{-4}$) and (SpT=M2, $F_{NUV}/F_J=5 \times 10^{-5}$). For M2--M4 stars, the lower envelope is defined by $F_{NUV}/F_J > 5 \times 10^{-5}$.

For stars later than M4, strong stellar activity can persist for a significant fraction of a Hubble time \citep[][]{West:2011zl}, limiting the usefulness of \galex\ data. However, for stars with SpT $<$ M4, these criteria would only reject 3 of our newly-identified Tuc-Hor members (subject to the spatial completeness of GALEX), which could themselves be field interlopers that are comoving by chance. We suggest that the optimal strategy for completing the Tuc-Hor census would be to use GALEX selection to identify the spatial distribution of Tuc-Hor members and remove most contaminants, and then to use kinematic+CMD selection to achieve the highest possible completeness within the spatial locus of Tuc-Hor members.

\acknowledgements

The authors thank Jason Curtis for obtaining many excellent observations as part of a time trade, Jason Wright for useful suggestions regarding the optimal map projection for plotting stars on the celestial sphere, and the anonymous referee for a helpful and thorough critique of the paper. ALK was supported in part by a Clay fellowship.

\clearpage

\bibliography{ms.bbl}

\clearpage

\LongTables

\clearpage
\begin{landscape}
\clearpage



\clearpage

\end{document}